\documentclass[onecolumn]{gji}
\let\bibhang\relax
\usepackage{natbib}
\usepackage{timet}
\usepackage{amsmath}

\usepackage{graphicx}
\usepackage{amssymb}
\usepackage{soul}

\usepackage{color}

\usepackage{bm}
\usepackage{verbatim}
\usepackage{textcomp}
\usepackage{bbding}
\usepackage{pifont}
\usepackage{mathrsfs}
\usepackage{yfonts}

\newcommand{\bx}{{\bf x}}

\newcommand{\bnabla}{{\boldsymbol{\nabla}}}

\newcommand{\cc}{{\mathcal C}}
\newcommand{\sgn}{{\mathrm{sgn}}}

\newcommand{\RR}{I\!\!R}

\newcommand{\rd}{\textrm{d}}
\def \i {\textrm{i}}
\def \r {\textfrak{r}}

\def \CC {\mathcal{C}}
\def\newblock{\hskip .11em plus .33em minus .07em}

\title[Interpreting Cross-correlations of One-bit Filtered Seismic Noise]{Interpreting Cross-correlations of One-bit Filtered Seismic Noise}
\author[Hanasoge \& Branicki]
  {Shravan M. Hanasoge$^{1,2}$ \& Micha\l\ Branicki$^3$ \\
  $^1$Department of Geosciences, Princeton University, NJ 08544, USA\\
  $^2$Max-Planck-Institut f\"{u}r Sonnensystemforschung, 37191 Katlenburg-Lindau, Germany.\\
  {Email: hanasoge@princeton.edu}\\
   $^3$Courant Institute for Mathematical Sciences, New York University, NY 10012, USA\\ 
  }
\date{Received 2012 June 18; in original form 2012 April 12}

\let\leqslant=\leq

\begin{document}

\label{firstpage}

\maketitle

\begin{summary}
Seismic noise, generated by oceanic microseisms and other sources, illuminates the crust 
in a manner different from tectonic sources, and therefore provides independent
information. The primary measurable is the two-point cross-correlation, evaluated using traces recorded at a pair of seismometers
over a finite-time interval.
However, raw seismic traces contain intermittent large-amplitude perturbations arising from tectonic activity and instrumental errors,
which may corrupt the estimated cross-correlations of microseismic fluctuations.
In order to diminish the impact of these perturbations, the recorded traces are
filtered using the nonlinear one-bit digitizer, which replaces
the measurement by its sign. 
Previous theory shows that for stationary Gaussian-distributed seismic noise fluctuations 
one-bit and raw correlation functions are related by a simple invertible transformation. 
Here we extend this to show that the simple correspondence
between these two correlation techniques remains valid for {\it non-stationary} Gaussian and a very broad range of {\it non-Gaussian}
processes as well.
For a limited range of stationary and non-stationary Gaussian fluctuations, we find that one-bit filtering performs at least as well as
spectral whitening.
We therefore recommend using one-bit filtering when processing terrestrial seismic noise, with the substantial benefit
that the measurements are fully compatible with current theoretical interpretation (e.g., adjoint theory). 
Given that seismic records are non-stationary and
comprise small-amplitude fluctuations and intermittent, large-amplitude
tectonic/other perturbations, we outline an algorithm to accurately retrieve the correlation function of the small-amplitude signals.
\end{summary}

\begin{keywords}
Theoretical Seismology -- Wave scattering and diffraction -- Wave propagation.
\end{keywords}

\section{Introduction}
The study of terrestrial seismic noise correlations holds great promise as a means of generating new crustal constraints 
and studying its temporal variations \citep[e.g.,][]{campillo07, wegler07,BrenguierSCI08,shapiro11, rivet11}.
These low amplitude waves are excited by storms, oceanic wave microseisms and related mechanisms 
\citep[e.g.,][]{longuet50,nawa98, kedar05, stehly06} at a variety of frequencies.
Because these sources are typically not co-spatial with tectonically active regions, waves thus created illuminate
the crust differently. Further, these types of seismic sources constantly excite wave noise, implying a continuous
ability to monitor the crust.

A widely used technique in the processing of continuous seismic noise measurements \citep[discovered by][]{weichert04} consists of applying a digitizing
{\it one-bit} filtering, introduced to the field first by \citet{aki65}. The method is straightforward, involving replacing
the raw measurement by its sign (after removing secular variations), certainly a very useful
technique in ca. 1965 when computer memory was highly limited \citep[which is why it was used; private communication,
G. Ekstr\"{o}m 2012; also see ][]{fink99, larose04, tanimoto07}. 
\cite{cupillard11} also studied this problem and derived a relationship which connects the raw cross-correlation
to the one-bit correlation. However \cite{cupillard11} make some incorrect assumptions in their analysis, leading them to
a result that differs from those of \cite{vanvleck} and others.

The method has gained traction over subsequent decades and is used very widely in seismic noise 
analysis today. The primary benefits attributed to one-bit digitized over raw 
processing are the superior stability of correlation measurements.
Observers who actively deal with such measurements undoubtedly `know' \citep[for instance,][]{fink99,larose04} that finite-time estimates of correlation functions obtained from sufficiently long one-bit digitized signals
tend to provide accurate and robust approximations of `true' correlations. Here, we approach this problem more systematically and shed some light on the reasons behind the success of this method in seismological applications.
Throughout this work, we assume that  the `true' signal arises from noise sources with well behaved statistics, i.e.,
we consider these sources with a bounded variance. In such a setting, we first discuss desirable properties of one-bit and other non-linear modulators in the idealized infinite-time case. In particular, we show that  the normalized one-bit and raw correlation functions are approximately related by a simple trigonometric transformation; this relationship is exact for a Gaussian seismic noise and a wide class of non-Gaussian perturbations. Insight into the finite-time artifacts and advantages of the one-bit  correlation techniques  are studied subsequently by corrupting the synthetic small-amplitude seismic noise traces with intermittent large-amplitude (tectonic) spikes. 

 We show in the following sections that one-bit clipping of the measured seismic signals works extremely well in diminishing the impact of intermittent large-amplitude events especially when estimating the correlations from a finite-time sample. Further, it is believed that the one-bit correlation is a better approximation of Green's function between the stations \cite[e.g.,][]{larose04}.
However, we show here that it is merely a better measurement of the noise cross-correlation when these large spikes are present in the traces \citep[note that Green's function and cross-correlations
are in general different, e.g., ][]{hanasoge12_sources}. 

The study of correlations of
digitized stochastic Gaussian-distributed processes has a long history, stretching back almost 80 years. In particular, we draw
from two articles, \citet{vanvleck} and \citet{hall}, in which the former calculate the correlation function of digitized Gaussian processes
while the latter estimate the signal-to-noise degradation incurred due to the application of the one-bit filter. 
In the field of seismology, it was discussed by e.g., \citet{tomoda56,aki57}, but is not used in contemporary
noise tomography and its properties have not been carefully investigated for relevant problems.

We reproduce some aspects 
of that calculation in the context of a digitized bivariate Gaussian process; this is done first within a simple Gaussian framework and, subsequently, 
as a particular case for non-Gaussian statistics. These results 
are then validated and illustrated on the 2-D models developed in \citet{hanasoge12_sources}. 
We study configurations of stationary and non-stationary sources to characterize the performance of the one-bit correlation. 
Finally, we examine the impact of `earthquake' type perturbations on the estimated correlations of the seismic noise. 
In all cases, we find that the one-bit correlation method performs at least as well as when correlating raw measurements, and is 
markedly superior (i.e., in that it is closer to the `true' noise correlation) when `earthquakes' are present. In the appendix, we 
consider the generalized framework for computing the correlation functions of the output for a large family of non-linear filters applied to a range of non-Gaussian stochastic processes.

\section{Statistics of Southern-California noise measurements}
\begin{figure}
\centering
\includegraphics*[width=\linewidth]{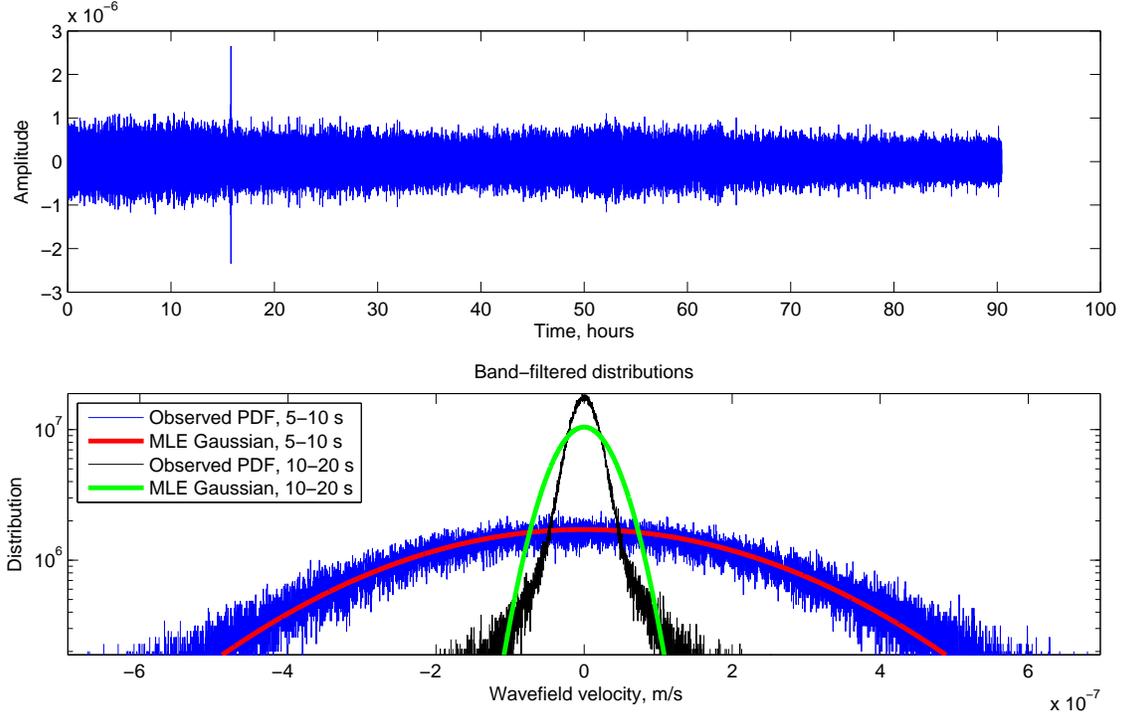}
\caption{A 68-hour long record of the vertical seismic velocity trace measured at station `ADO' in Southern California (upper panel)
and the probability density function (pdf) shown in the lower panel. A maximum-likelihood Gaussian fit is shown in the red curve in the lower panel.
Although the trace (upper panel) shows large-amplitude spikes, a Gaussian well approximates the empirical PDF in the 5-10 second band
with $D_{\rm KL}= 0.026$. In comparison, the 10-20 second band whose $D_{\rm KL} = 0.25$ is more non-Gaussian. How then do we process data that have
large spikes, which are likely to significantly bias the correlations? How do we diminish the impact of the spikes while still 
extracting the information present in the noise? 
\label{pdf_socal}}
\end{figure}

The underpinning of the theoretical arguments we lay out here is the assumption that seismic noise fluctuations are
Gaussian random processes. To demonstrate that this is a valid assumption, we study measurements of vertical seismic velocity taken by
station `ADO', operated by California Institute of Technology, in the tectonically active region of southern California. 
We use 68 hours of measurements, taken during Jan 1-3 2010, which are frequency-filtered in the ranges 5-10 and 10-20 seconds. From this time series, we
construct a histogram of the fluctuations and normalize it to obtain the empirical probability density function (pdf) 
of vertical seismic velocity. The data and the pdf are plotted
in Figure~\ref{pdf_socal}. 

\begin{figure}
\centering
\includegraphics*[width=\linewidth]{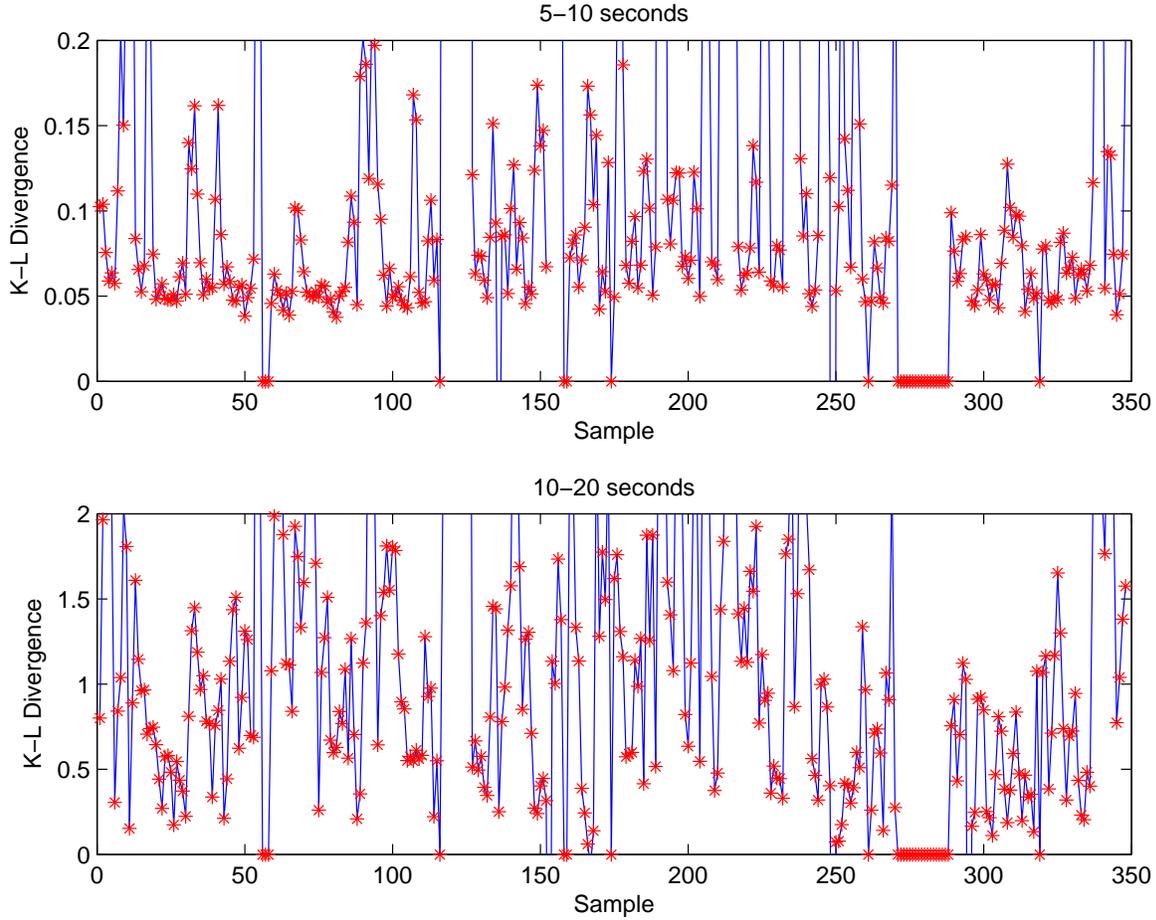}
\caption{Relative entropy $D_{\rm KL}(P \| Q)$ between seismic noise measurements and maximum-likelihood Gaussian fits using
the Kullback-Leibler (K-L) divergence (Eq.~[\ref{kldivform}]) as the metric. We take 348 68-hour sequences of seismic noise
measurements from station `ADO' and filter these data to isolate the 5-10 and 10-20 second period oscillations. Some 31 of these samples have
measurement problems, leaving us with 317 useful windows.
We then perform maximum-likelihood
Gaussian fits to each sequence and measure the K-L divergence between the fit and the measurement. There are problems with 
the data (such as truly massive, unrealistic spikes), and these cause the K-L divergence to be very large. At the 5-10 second range For 265 of the 319 samples (83\%),
the value of the K-L divergence is less than 0.1, with 99.7\% of these samples possessing a K-L divergence of less than 0.2,  supporting our assumption of Gaussianity for oscillations in this frequency band. At the 10-20 second range, the data are less easily fit by a Gaussian, with 90\% of all samples possessing values of K-L divergence less than 2.
For a much more detailed analysis of the statistics of noise fluctuations, please see, e.g., \cite{groos2009}.
\label{kldiv}}
\end{figure}

Assuming the data are Gaussian-distributed, we perform a maximum-likelihood fit
(shown in red, lower panel). We compare the two pdfs using the Kullback-Leibler (K-L) divergence,
which is an information-theoretic measure of the distance between a pair of pdfs. In the discrete case, the K-L divergence is given by,
\begin{equation}
D_{\rm KL}(P \| Q) = \sum_i P(i)\,\ln\frac{P(i)}{Q(i)},\label{kldivform}
\end{equation}
where $P$ is the empirical (observed) pdf and $Q$ is the fit. We choose a Gaussian model for $Q$ to perform the fit. 
The smaller the value of the K-L divergence, the better
the information contained in a process with pdf $P$ is captured by the model $Q$; the K-L divergence is zero only when $P = Q$.
Figure~\ref{pdf_socal} shows empirical pdfs and corresponding maximum-likelihood fits associated with 5-10 and 10-20 second bands 
for a 68-hour sequence. In Figure~\ref{kldiv},
we choose 348 instances of 68-hour measurement windows, perform a maximum-likelihood 
fit a Gaussian pdf to each interval and measure the K-L divergence between the data and the fit.
A non-zero K-L divergence indicates non-Gaussianty. In general, non-Gaussianity may take arbitrary
forms, i.e., long tails, oscillations in the distribution around a Gaussian etc. and no metric can unambiguously quantify the nature of the departure
from a Gaussian distribution (i.e., non-Gaussianity is a non-unique notion). One can therefore only compare K-L divergences to
say that the 5-10 second band is much better fit by a Gaussian than the 10-20 second series (see Figures~\ref{pdf_socal} and~\ref{kldiv}).
A much more detailed picture of the distribution and stationarity characteristics of seismic fluctuations may be found in \cite{groos2009}.
Ultimately, neither Gaussianity nor stationarity are necessary for one-bit processing since we have developed a theory to address seismic fluctuations
drawn from a broad range of distributions.

\section{One-bit correlation}\label{gauss.one}
Guided by the preceding section, we model seismic noise in the 5-10 second bands
as a multivariate Gaussian random process. 
Consider two stationary correlated zero-mean random sequences $\{S_1(t)\}$ and $\{S_2(t)\}$, which represent seismic noise displacements measured
at stations 1 and 2. We 
are interested in the connection between one-bit-digitzed and raw cross-correlations between these two sequences. The one-bit filter behaves as a non-linear
Signum or sign function of the raw signal, mapping analogue signals, defined on the set of real numbers, to a digital $[-1, 1]$:
\begin{equation}
\sgn(x) = 
\begin{cases} 1 & \text{if $x \ge 0$,}
\\
-1 &\text{if $x< 0$.}
\end{cases}
\end{equation}
The finite-time estimate of the cross-correlation function of zero-centered seismic noise fluctuations recorded at two stations, $s_1(t), s_2(t)$, denoted by $\tilde\cc(t)$, is defined as
\begin{equation}
\tilde\cc(\tau) = \frac{1}{T}\int_0^T\,dt\,s_1(t)\,s_2(t+\tau),\label{cctime}
\end{equation}
where $T$ is the temporal length of averaging and $t$ is time. We note that the normalized finite-time correlation function $\tilde\rho(\tau)$ is related to the cross-correlation through \begin{equation}
\tilde\rho(\tau) =  \frac{\tilde\cc(\tau)}{\tilde\sigma_1\tilde\sigma_2},\label{corr.func}
\end{equation}
where the finite-time variance estimates of $s_1$ and $s_2$ are defined analogously as 
\begin{equation}
\tilde \sigma_i^2 = \frac{1}{T}\int_0^Ts_i^2(t)  \,\rd t.
\end{equation}
If the processes $s_i$ are ergodic and have bounded non-degenerate covariance (which we implicitly assume throughout) the cross-correlation and the correlation function are given, respectively, by 
\begin{equation}\label{cinf}
a) \quad C(\tau)\equiv \langle s_1(t)\,s_2(t+\tau)\rangle =  \underset{T\rightarrow \infty}{\lim} \tilde \cc(\tau),
\hspace{1.5cm} 
b) \quad \rho(\tau) =  \underset{T\rightarrow \infty}{\lim} \tilde \rho(\tau),\hspace{1.5cm}
c) \quad \sigma_i =  \underset{T\rightarrow \infty}{\lim} \tilde \sigma_i,
\end{equation}
where $\langle \cdot\rangle$ denotes the statistical ensemble average and the existence of the limits follows from the ergodicity assumption.
In practice, the length of the averaging time interval $T$ must be at least several source correlation times or inter-station wave travel times (whichever is larger). 
If the noise fluctuations were an ergodic random process with bounded variance, it follows that the finite-time estimate of the cross-correlation function $\tilde{\mathcal{C}}$ in (\ref{cctime}) approaches the true cross-correlation $\mathcal{C}$ in (\ref{cinf}a) as the temporal window of averaging $T$ grows. 

The primary goal of this section is to elucidate the link between  the (infinite-time) cross-correlation of the input $\cc$ in (\ref{cinf}) to the cross-correlation $\cc^1$ of the output of the one bit digitizer  given by 
\begin{equation}
\cc^1(\tau) = \big\langle \sgn(s_1(t))\, \sgn(s_2(t+\tau))\big\rangle,
\end{equation}
with the normalized correlation function defined analogously to (\ref{cinf}b) and denoted by $\rho^1(\tau)$.
An elegant derivation of the link between the correlation function of the true Gaussian signal and the output of the one-bit filter was given in \cite{vanvleck}; we briefly recapitulate it below in a suitable `bivariate' formulation since this classical results has not been discussed in noise tomography applications. An alternative derivation of this result in a much more general framework is presented in Appendix \ref{gen_form}. 
Measurements at two stations may be treated as a bivariate Gaussian process,
with correlation matrix $\Sigma(\tau)$,
\begin{eqnarray}
f(S_1,S_2,\tau) &=& \frac{1}{2\pi \sqrt{|\Sigma|}}\exp\left(-\frac{1}{2}{\bf S}^T\,\Sigma^{-1}\,{\bf S}\right),\nonumber\\
&=& \frac{1}{2\pi\sigma_1\sigma_2\sqrt{1-\rho(\tau)^2}}\exp\left[-\frac{1}{2(1-\rho(\tau)^2)}\left(\frac{S_1^2}{\sigma_1^2}
+\frac{S_2^2}{\sigma_2^2} - 2\rho(\tau) \frac{S_1 S_2}{\sigma_1\sigma_2}\right)\right],
\end{eqnarray}
where ${\bf S} = [S_1\,\, S_2]^T$ and $|\Sigma|$ is the determinant of the matrix. The correlation function of the two random variables $S_1,S_2$
given their joint pdf is defined as
\begin{equation}
\cc(\tau) =  \int_{-\infty}^\infty dS_1\int_{-\infty}^\infty dS_2\,\,\, S_1\,S_2\,\,f(S_1,S_2,\tau).
\end{equation}
Thus we have,
\begin{equation}
\cc(\tau) = \int_{-\infty}^\infty dS_1\int_{-\infty}^\infty dS_2\,\,\, S_1\,S_2\,\frac{1}{2\pi\sigma_1\sigma_2\sqrt{1-\rho(\tau)^2}}\exp\left[-\frac{1}{2(1-\rho(\tau)^2)}\left(\frac{S_1^2}{\sigma_1^2}
+\frac{S_2^2}{\sigma_2^2} - 2\rho(\tau) \frac{S_1 S_2}{\sigma_1\sigma_2}\right)\right],
\end{equation}
and upon defining $X = S_1/\sigma_1$ and $Y = S_2/\sigma_2$, the integral reduces to
\begin{equation}
\frac{\cc(\tau)}{\sigma_1\sigma_2} = \int_{-\infty}^\infty dX\,\int_{-\infty}^\infty dY\,\, X\,Y\,\frac{1}{2\pi\sqrt{1-\rho(\tau)^2}}\exp\left[-\frac{1}{2(1-\rho(\tau)^2)}\left(X^2
+Y^2 - 2\rho(\tau) XY\right)\right].\label{corr.def}
\end{equation}
Comparing with equation~(\ref{corr.func}), we note that the left side of equation~(\ref{corr.def}) is $\rho(\tau)$.
The correlation function of one-bit filtered random variables $X, Y$ is therefore given by
\begin{equation}
\rho^1(\tau)= \int_{-\infty}^\infty dX\,\int_{-\infty}^\infty dY\,\, \sgn(X)\,\sgn(Y)\,\frac{1}{2\pi\sqrt{1-\rho(\tau)^2}}\exp\left[-\frac{1}{2(1-\rho(\tau)^2)}\left(X^2
+Y^2 - 2\rho(\tau) XY\right)\right].
\end{equation}
This integral may be evaluated in closed form
\begin{equation}
\rho^1(\tau)= \frac{1}{2\pi\sqrt{1-\rho(\tau)^2}}\left[\int_{0}^\infty \int_{0}^\infty e^{-\chi}dX\, dY
- \int_{-\infty}^0 \int_{0}^\infty e^{-\chi}dX\, dY - \int_{0}^\infty \int_{-\infty}^0 e^{-\chi}dX\, dY + \int_{-\infty}^0 \int_{-\infty}^0 e^{-\chi}dX\, dY\right],\label{split1}
\end{equation}
where
\begin{equation}
\chi \equiv\frac{1}{2(1-\rho(\tau)^2)}\left(X^2+Y^2 - 2\rho(\tau) XY\right).
\end{equation}
Using the identity that the total integral of the probability density function is 1, i.e., 
\begin{equation}
\frac{1}{2\pi\sqrt{1-\rho(\tau)^2}}\int_{-\infty}^\infty dX\,\int_{-\infty}^\infty dY\,e^{-\chi} = 1,
\end{equation}
we may rewrite the split in equation~(\ref{split1}) as
\begin{equation}
\rho^1(\tau)= \frac{4}{2\pi\sqrt{1-\rho(\tau)^2}}\,\int_{0}^\infty \int_{0}^\infty e^{-\chi}dX\, dY - 1.
\end{equation}
We introduce the transform $X = r\cos\phi, Y = r\sin\phi$, allowing us to rewrite the integral as
\begin{equation}
\rho^1(\tau)= \frac{4}{2\pi\sqrt{1-\rho(\tau)^2}}\left\{\int_{0}^{\pi/2} \int_{0}^\infty \exp\left[-\frac{r^2}{2(1-\rho(\tau)^2)}\left(1 - \rho(\tau) \sin 2\phi\right)\right]rdr\, d\phi\right\} -1,\label{split2}
\end{equation}
where the integral in (\ref{split2}) may be evaluated using the following identity
\begin{equation}
\int_{0}^\infty e^{-\alpha r^2}\,rdr = -\frac{1}{2\alpha}\int_{0}^\infty\, dr\,\partial_r(e^{-\alpha r^2}) = \frac{1}{2\alpha}.
\end{equation}
Consequently, the correlation functions $\rho^1(\tau)$ may be written as 
\begin{equation}
\rho^1(\tau)= \frac{2\sqrt{1-\rho(\tau)^2}}{\pi}\,\int_{0}^{\pi/2} \frac{d\phi}{1-\rho\sin 2\phi} -1,
\end{equation}
where the integral above may be evaluated using the well-known formula 
\begin{equation}
\int_0^{\pi/2} \frac{dx}{a + b \cos x} = \frac{\arccos(b/a)}{\sqrt{a^2 - b^2}}.
\end{equation}
The transfer function between one bit to raw correlations is given by
\begin{equation}
\rho^1(\tau) = \frac{2}{\pi}\arcsin\rho(\tau),
\end{equation}
or, conversely,
\begin{equation}
\rho(\tau) = \sin\left(\frac{\pi}{2}\rho^1(\tau)\right).\label{onetorawinfinite}
\end{equation}
Thus, for two stationary Gaussian processes there exists a simple one-to-one mapping (\ref{onetorawinfinite}) between the correlation function $\rho(\tau)$ obtained from the raw data and the digitized one-bit correlation function $\rho^1(\tau)$; see Appendix \ref{gen_form} for a derivation of analogous result in a much more general framework.
In what follows, we assume that the relationship (\ref{onetorawinfinite}) holds approximately for the finite-time estimates, $\tilde \rho$ and $\tilde \rho^1$, of the normalized raw data and one-bit cross-correlation functions, i.e., 
\begin{equation}
\tilde\rho(\tau) = \sin\left(\frac{\pi}{2}\tilde\rho^1(\tau)\right).\label{onetoraw}
\end{equation}
We omit here the technical justification of the above assumption for brevity; intuitively, this can be seen by noticing that for two stationary ergodic processes, $s_1$ and $s_2$, with finite means and variances the finite-time estimates of the correlation functions $\tilde\rho(\tau),\; \tilde\rho^1(\tau)$  are Gaussian random variables with respective means given by $\langle \tilde\rho(\tau)\rangle =\rho(\tau),\;\langle \tilde \rho^1(\tau)\rangle= \rho^1(\tau)$ and variances bounded by $Var[\tilde\rho(\tau)]\leqslant Var[\tilde\rho(\tau)]/T$ and $Var[\tilde\rho^1(\tau)]\leqslant Var[\tilde\rho^1(\tau)]/T$ for any $T\ne 0$ in (\ref{cctime}). 

\medskip
A few remarks are in order here. First, the relationship (\ref{onetorawinfinite}) is strictly valid for stationary Gaussian signals; a more general formula which does not require stationarity is derived in Appendix \ref{gen_form} (see (\ref{CC_gss_nonst})). Second, the both the exact formula in (\ref{onetorawinfinite}) and its finite-time approximation (\ref{onetoraw}) assume that the seismic noise signal is not corrupted by other sources. However, in the remainder of this paper we show that the recovery of the estimates of the correlation function $\rho(\tau)$ from the one-bit digitized signal via (\ref{onetoraw}) provides superior results to those obtained from finite time estimates, $\tilde \rho(\tau)$, obtained directly from the raw data. This is illustrated using both the numerical experiments and analytical results discussed in Appendix \ref{gss_ngss} where we  show that the relationship between the correlation function of the one-bit digitized signal $\rho^1(\tau)$ and the correlation function of the Gaussian seismic noise $\rho(\tau)$ is given, to a very good approximation, by the  simple trigonometric formula (\ref{onetoraw}) established  in the Gaussian context. Remarkably,  for a large class of bivariate non-Gaussian processes corrupting the accuracy of (\ref{onetoraw}) is insensitive to the correlations between the corrupting processes, as also shown in Appendix \ref{gss_ngss}.

\subsection{The transfer function and signal-to-noise ratio}
\begin{figure}
\centering
\includegraphics*[width=\linewidth]{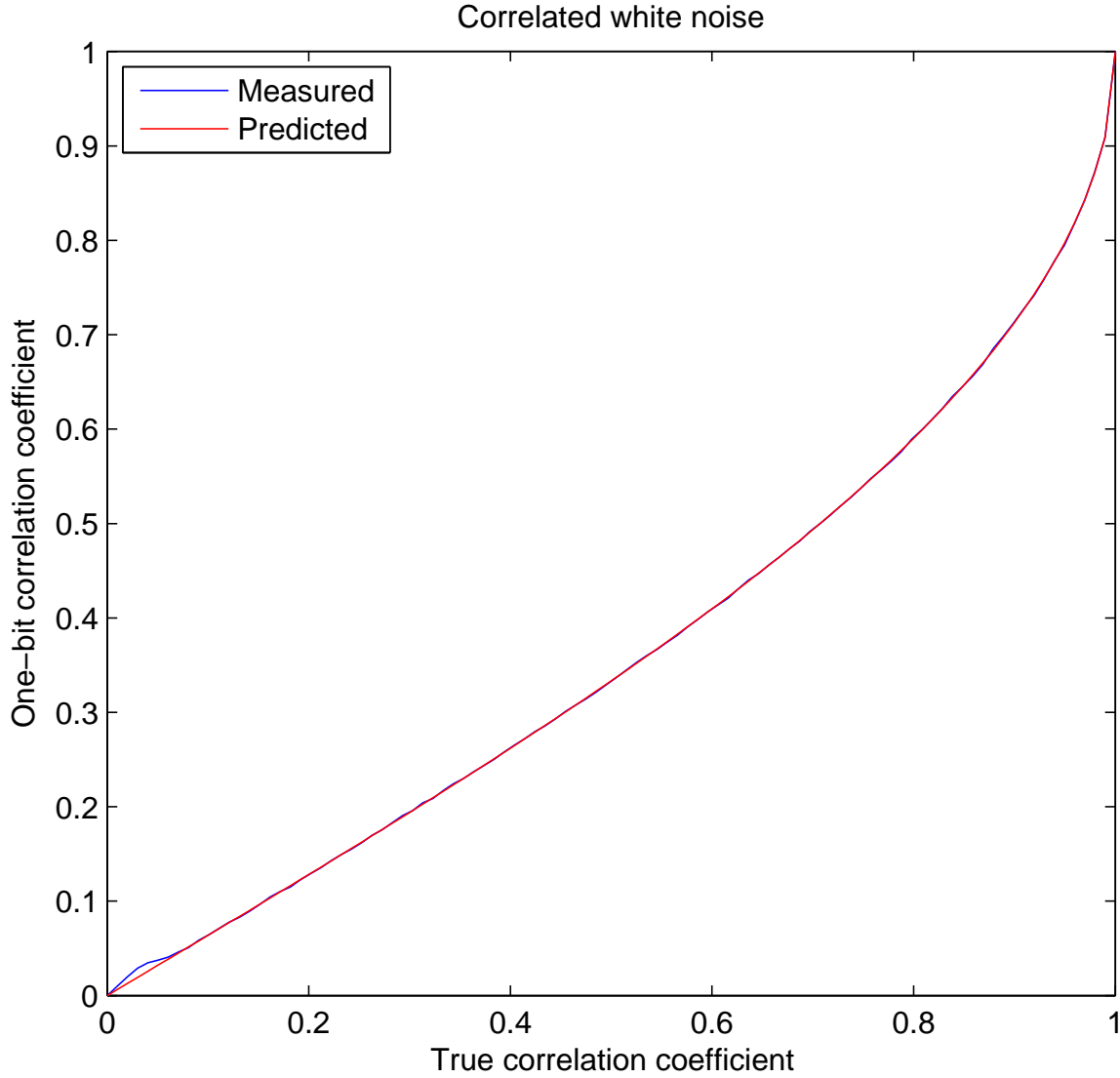}
\caption{True and one-bit cross-correlation coefficients of sequences of a bivariate Gaussian random variable.
The stochastic processes are white noise and delta-correlated, implying that the correlation coefficient away from zero-time lag is statistically zero.
We vary the true correlation coefficient ($x$ axis) while keeping the standard deviations fixed.
The corresponding predicted and measured one-bit coefficients are plotted on the $y$ axis. We confirm numerically
that equation~(\ref{onetoraw}) is satisfied.
\label{corrcoeff}}
\end{figure}

To numerically confirm equation~(\ref{onetoraw}), we generate sequences of delta-correlated random variables, i.e., where the correlation
coefficient away from zero time-lag is statistically zero. This implies that there is only one characteristic correlation coefficient. It therefore allows us
to easily test the accuracy of the transfer function~(\ref{onetoraw}), by comparing the one-bit-filtered and true correlation coefficients in Figure~\ref{corrcoeff}.

Because the digitizing filter results in
the loss of phase information, the one bit {output} is noisier than the raw correlations. 
Evidently, the degradation in signal-to-noise ratio is dependent on a variety of factors, most importantly the specific definition of SNR and 
related issues such as the
amount of noise present in the traces. 
\subsection{Band-limited Gaussian Random variables}
Relation~(\ref{onetoraw}) provides us with a measurement algorithm: 
\begin{itemize}
\item Apply digitizing filter on raw seismic traces,
\item Compute cross-correlations, 
\item compute normalized correlation coefficient (Eq.~[\ref{corr.func}]) and 
\item apply transfer function~(\ref{onetoraw}).
\end{itemize}

\begin{figure}
\centering
\includegraphics*[width=0.75\linewidth]{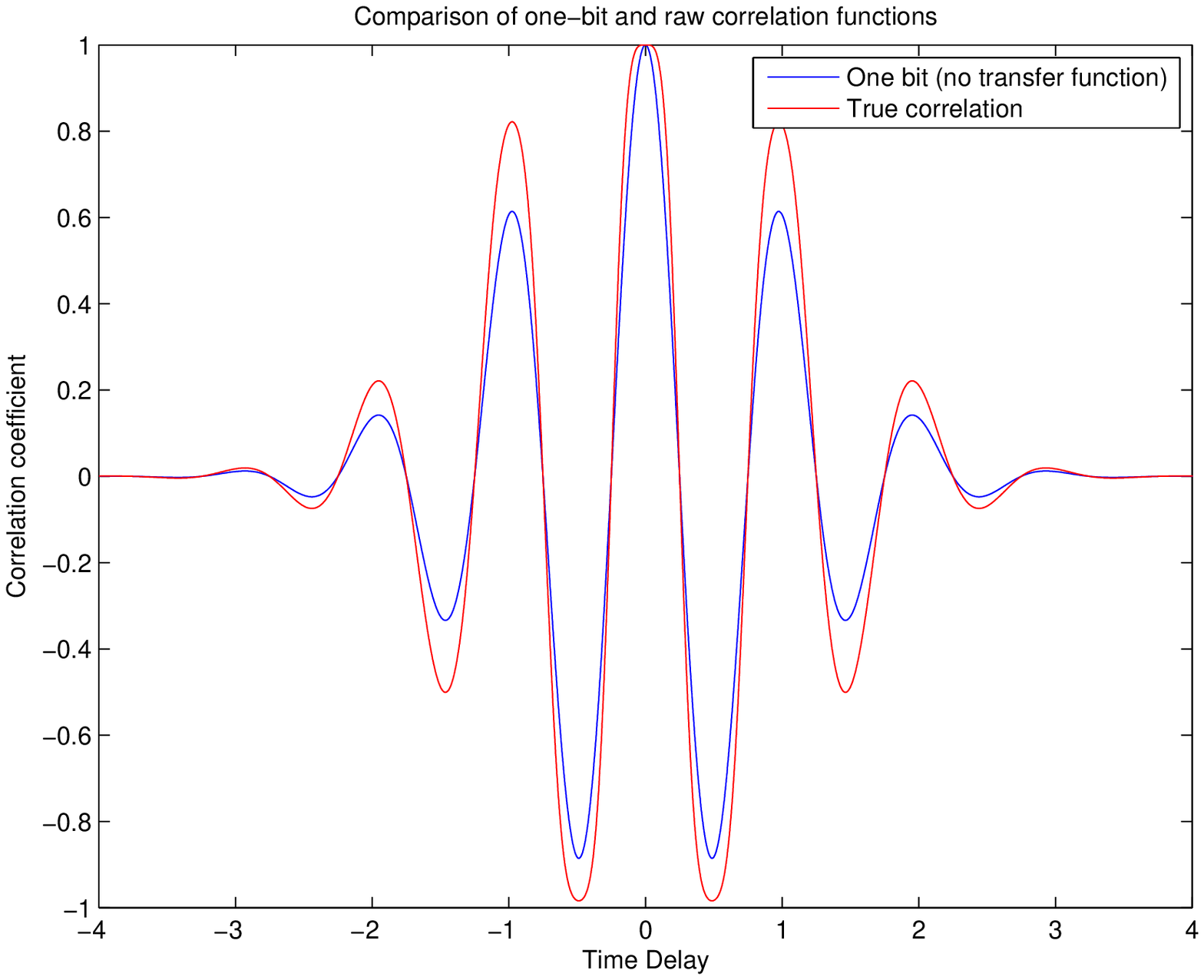}
\caption{The difference between the correlations of raw and one-bit filtered variables (when, as in current practice, no transfer function is applied).
An artificially generated `one-bit' correlation function, $\rho^1(\tau)$, (blue) which upon application of transfer function~(\ref{onetorawinfinite})
gives us $\rho(\tau)$ (red), the `true' correlation. The two functions are similar, explaining why
efforts in the past that have used one-bit correlations have generally resulted in seemingly stable and rational results. However, they are different
and to improve the precision and accuracy of measurements, it is necessary to use transfer function~(\ref{onetoraw}).
Unless specified explicitly, the output of this algorithm will be termed `One-bit correlation' in subsequent figures.
 \label{onerawcomp}}
\end{figure}

Terrestrial seismic noise is temporally band limited (owing, in large part, to instrumental limitations) and so it is natural to study this case next.
In Figure~\ref{onerawcomp}, we show an artificially generated one-bit correlation function, i.e., $\rho^1(\tau)$, (blue) and
the `true' correlation function, which is defined by transfer function~(\ref{onetorawinfinite}), i.e., $\rho(\tau)$. It is seen
that the functions are similar but to improve the precision and accuracy of the measurement, it is necessary to apply relation~(\ref{onetorawinfinite}).
Unless specified explicitly, the output of this algorithm, i.e., where we have applied transfer function~(\ref{onetoraw}), will be termed `One-bit correlation' in subsequent analyses and figures.

\begin{figure}
\centering
\includegraphics*[width=\linewidth]{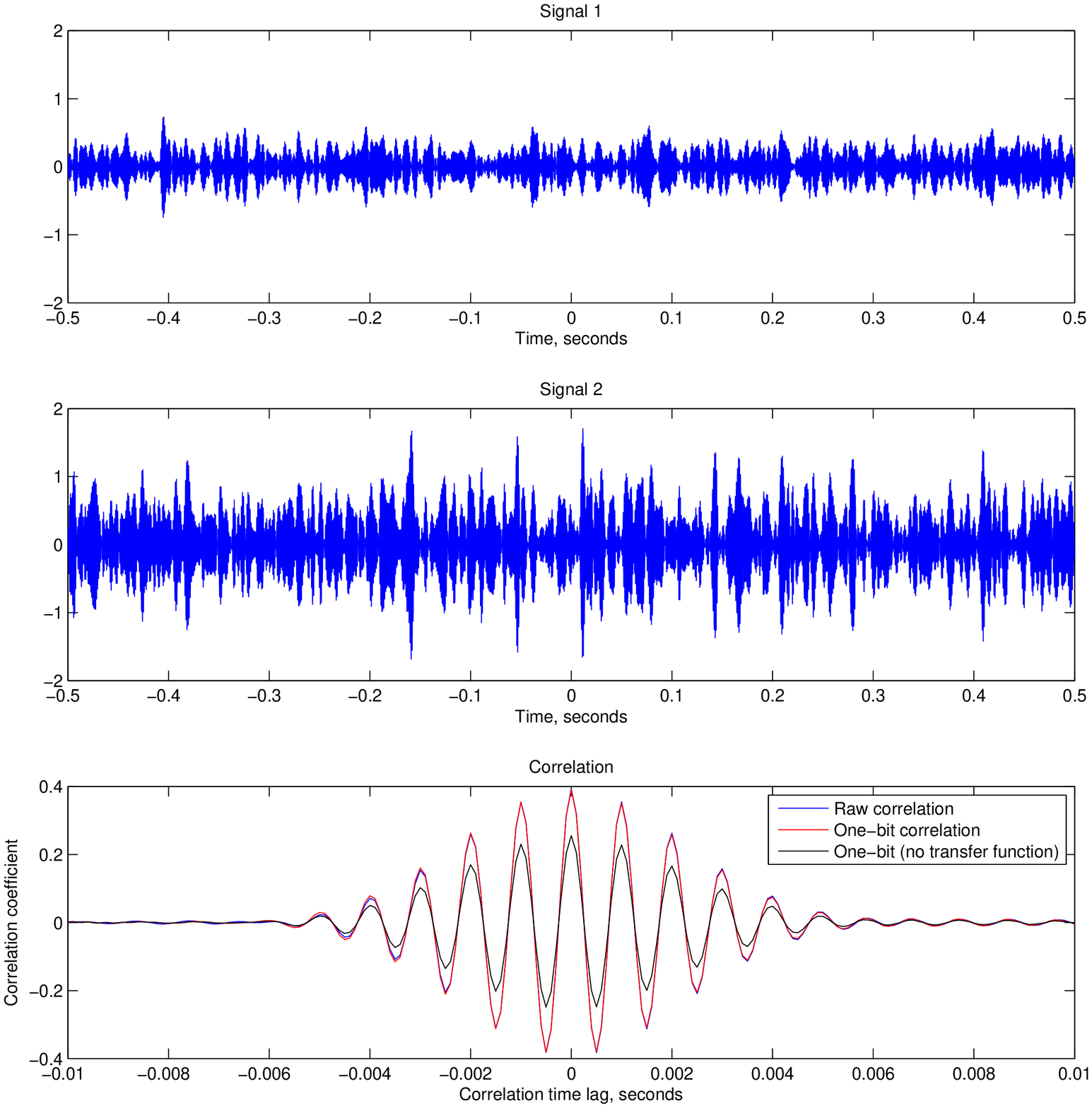}\vspace{-1cm}
\caption{Raw and one-bit cross-correlations of sequences of a band-limited bivariate Gaussian random variable.
The correlation of the one-bit filtered traces with and without having applied transfer relation~(\ref{onetoraw}) are shown as well.
Once the transfer function is applied, one-bit and raw correlations are indistinguishable (red and blue lines) whereas 
errors are incurred if the transformation is not performed (black line; as is current practice).
This is just an exercise in signal processing and no wave propagation physics is taken into account
at this stage.\label{gaussian_sp}}
\end{figure}

We confirm the statistical equivalence of one-bit and raw correlations of a stationary band-limited bivariate Gaussian random variable in Figure~\ref{gaussian_sp}.
Correlations, raw and one bit, of a large number of realizations of band-limited random variables are averaged to obtain estimates of the expectation value. It is seen that the agreement
between the two is excellent (having applied the transfer function~(\ref{onetoraw}) on the correlation of the one-bit filtered traces). 

In appendix~\ref{corrgen}, we extend the results to stationary random variables drawn from a broad range of distributions. 
Gaussianity is therefore not a necessary condition for the oscillations.
We also invoke stationarity, a central assumption in this analysis,
one that is not generally representative of measurements. For non-stationary oscillations, the theory becomes much more complicated
since the meaning of the ``true" cross-correlation is not clear.
For a stationary random variable with bounded variance, 
the one-bit correlation with transfer function has the {\it same} expectation value as the raw correlation. 

Similar to this work, \cite{cupillard11} developed a theory of one-bit filtering based on concepts of ``coherent" and ``incoherent" noise, using it to characterize the 
functional relationship between one-bit and raw correlations. Their formalism requires knowing
the coherent and incoherent constituents {\it a priori}, which may be very difficult to obtain.
In contrast, with no knowledge of the coherence or the lack thereof, we provide an algorithm to 
estimate the true correlation coefficient given the one-bit measurement.

The main result of \cite{cupillard11}, expressed in their equation (52), which connects the raw cross-correlation
to the coherent and incoherent parts of the one-bit correlation, is slightly erroneous. In their appendix C, \cite{cupillard11} calculate 
the probability of the correlation of the two one-bit filtered traces assuming they are independent, which is in error. This leads them to
a result that differs from those of \cite{vanvleck} and others.

\begin{figure}
\centering
\includegraphics*[width=\linewidth]{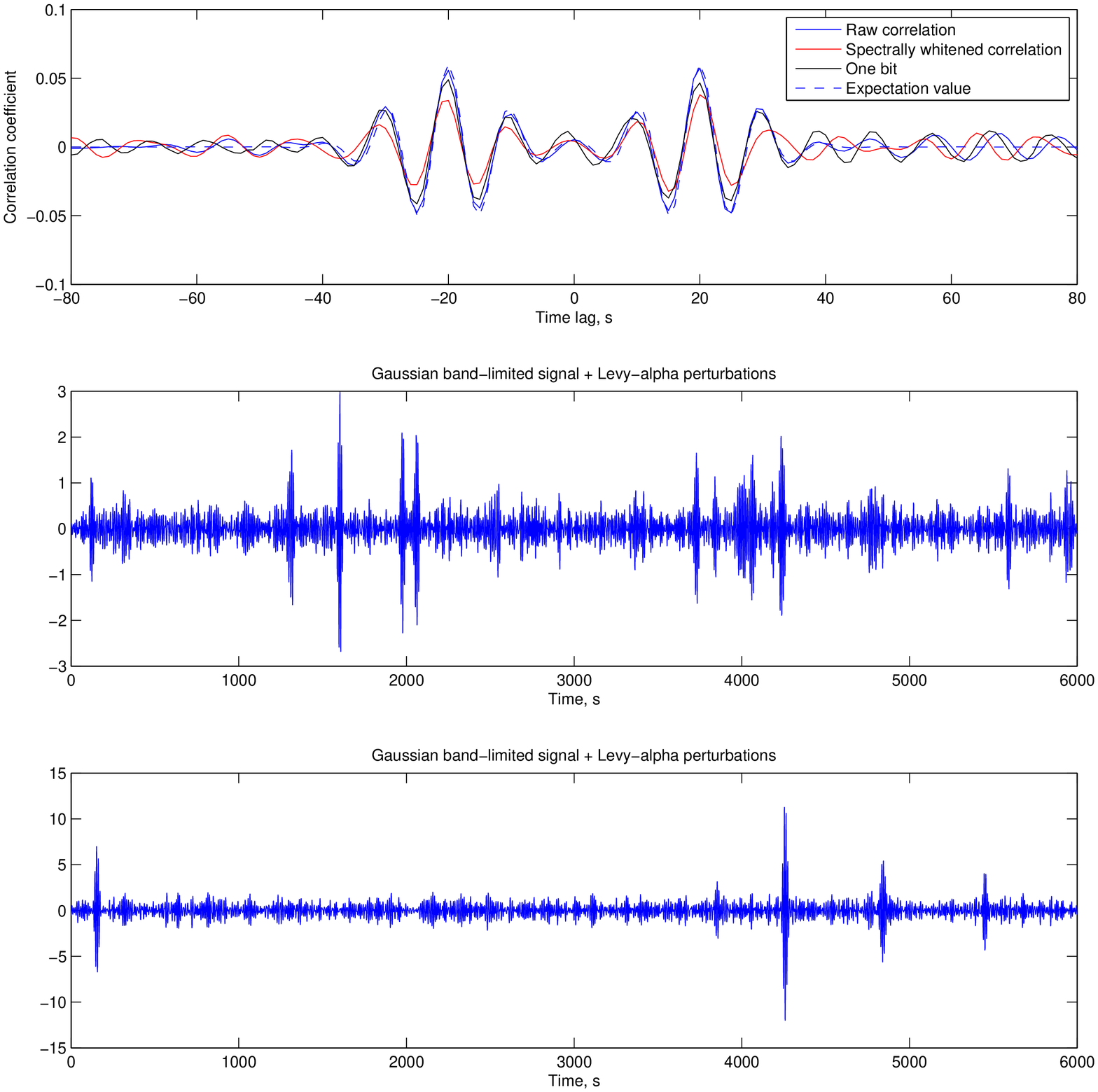}\vspace{-1cm}
\caption{A comparison between raw, one-bit and spectral whitening strategies \citep[we use a variant of the method suggested by][]{seats12} for bivariate Gaussian band-limited variables
with transient perturbations drawn from a L\'{e}vy stable distribution. It is seen that the one-bit and raw variables more closely match the 
expected value of the correlation (which is just the correlation function of the Gaussian variables). A realization of 6000 seconds was generated
to create this plot. For the spectral whitening measurement, this realization was split into segments of 300 seconds long and the correlations
of each segment were stacked together.  \label{whitening}}
\end{figure}

In Figure~\ref{whitening}, we compare the method of spectral whitening to one-bit and raw correlations. Spectral whitening
allows for a number of free parameters and our intention is not to explore the full regime; we assume the form~(\ref{whitespec}),
which is a variant of the method discussed by \citet{seats12}.
We compute the correlation of a stationary band-limited Gaussian random signal with a few transient perturbations drawn from a L\'{e}vy power-law distribution.
The 6000-second signal in Figure~\ref{whitening} is divided into segments of 200 seconds each, auto-correlations are computed for each
segment normalized by its power spectrum. These 30 correlations are averaged and multiplied by the average power spectrum of the segments.
In other words, the spectrally whitened correlation function $\mathcal{C}^{\rm w}$ is computed so
\begin{equation}
\mathcal{C}^{\rm w} = \frac{1}{N}\sum_{i=1}^N \frac{X^*_i(\omega) Y_i(\omega)}{|X_i(\omega)||Y_i(\omega)|}\,  \sqrt{{\mathcal P}_X(\omega)\,{\mathcal P}_Y(\omega)},\label{whitespec}
\end{equation}
where original signals $X(t)$ and $Y(t)$ are divided into $N$ segments $X_i$ and $Y_i$, and with each segment multiplied by a window function to
set it smoothly to zero at the edges. The average power spectrum of the processes are denoted by ${\mathcal P}_X = |X(\omega)|^2$ and ${\mathcal P}_Y = |Y(\omega)|^2$.
The raw and one-bit correlations are accurate and closely match the expected value but the spectrally whitened correlation 
shows slightly larger errors in amplitude and phase. Even in the case when there are no transient perturbations, i.e., where the signal is purely Gaussian,
spectral whitening converges more slowly to the expected correlation amplitude than one-bit and raw correlations. 
Indeed, we do not perform an exhaustive parametric study, and it may be that in some regimes, spectral whitening does better.
We also note that, most importantly, there is a theoretical basis for interpreting the one-bit correlations and that with transfer function~(\ref{onetoraw}),
we recover the expectation value of the correlation function.

The question of what happens when the distribution of seismic noise fluctuations is non-Gaussian is indeed
valid. In appendix~\ref{corrgen}, we discuss
more generalized distributions (i.e., non-Gaussian fluctuations) and address a range of non-linear modulators. For the cases we have studied,
where the fluctuations possessed exponential/ other heavy-tailed distributions, we find that relation~(\ref{onetoraw}) still remains very effective. 

\subsection{Non-stationary processes}
\begin{figure}
\centering
\includegraphics*[width=\linewidth]{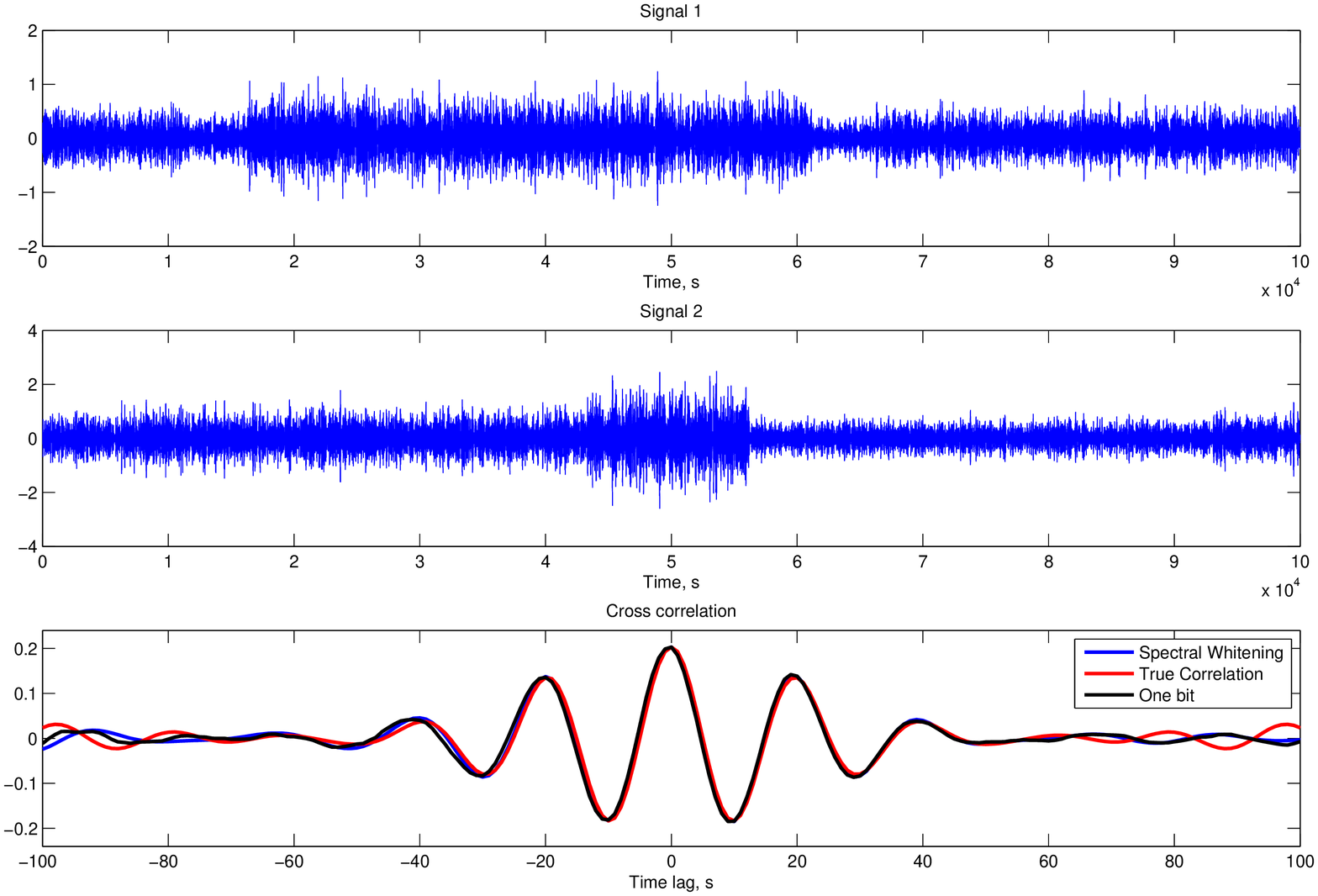}\vspace{-1cm}
\caption{A comparison between raw, one-bit and spectral whitening strategies \citep[we use a slightly different method from][]{seats12} for temporally non-stationary bivariate Gaussian band-limited variables. 
For the spectral whitening measurement, this realization was split into segments of 5000 seconds long and the correlations
of each segment were stacked together.  It is seen that the one-bit correlations and spectral whitening are indistinguishable.\label{nonstattemp}}
\end{figure}

We characterize the performance of spectral whitening and one-bit filtering applied to statistically non-stationary processes, a common feature
of observed seismic fluctuations. We generate sequences of stationary bivariate Gaussian signals and their correlation function is
termed the `true' correlation function. These processes are subsequently modulated in time so as to generate a process whose variance changes, and
thus is temporally non-stationary. 
From this test, we find in Figure~\ref{nonstattemp} that correlations derived from one-bit filtering are essentially identical to
those obtained by spectral whitening.

\section{Demonstration using the wave equation}
Here we study wave propagation in a 2-D homogeneous background, where wave excitation is effected by a temporally stochastic and spatially non-uniform
distribution of sources. The goal is to simulate wave noise, record this at a pair of stations and cross correlate them over
many realizations (or long periods of time; we assume ergodicity here). Raw and one-bit filtered noise traces are correlated and
compared to the (`true') expectation value.

The relation~(\ref{cross.c})
when applied to equation~(\ref{cctime}) gives us the cross-correlation function in the Fourier domain
\begin{equation}
\tilde\cc_{\alpha\beta}(\omega) = \phi^*(\bx_\alpha,\omega)\,\phi(\bx_\beta,\omega),\label{correl}
\end{equation}
where the $\phi$ are random variables and $\tilde\cc_{\alpha\beta}(\omega)$ is the Fourier transform
of the finite-time estimate of the cross-correlation.
Were it to exist, the limit (or expected) cross-correlation, denoted by $\cc_{\alpha\beta}(\omega)$, is
\begin{equation}
\cc_{\alpha\beta}(\omega) = \langle\phi^*(\bx_\alpha,\omega)\,\phi(\bx_\beta,\omega)\rangle.\label{expect}
\end{equation}
Note that $\tilde\cc_{\alpha\beta}(\omega)$ is a random variable whereas $\cc_{\alpha\beta}(\omega)$ is
a fixed quantity.

For the sake of simplicity, we consider wave propagation in a 2-D plane, described by
\begin{equation}
\rho\partial_t^2\phi - \bnabla\cdot(\rho c^2\bnabla\phi) = S(\bx,t),\label{2deq}
\end{equation}
where $\rho$ is density, $\bx= (x,y)$ is a 2-D flat space, $t$ time,
 $\phi$ the wave displacement, $\bnabla =( \partial_x, \partial_y)$ the covariant spatial derivative, $S(\bx,t)$ the
 source and $c$ wavespeed. We also assume a constant wavespeed $c$. 
 Green's function $G(\bx,\bx';t)$ for the displacement $\phi(\bx,t)$ due to a spatio-temporal delta source at $(\bx',0)$ is 
the solution to
 \begin{equation}
(\rho \partial^2_t - \rho c^2\nabla^2)G(\bx,\bx';t) = \delta(\bx-\bx')\delta(t).
 \end{equation}
Green's function in the Fourier domain for this wave equation is given by
 \begin{equation}
 G(\bx,\bx',\omega) = H_0^{(1)}\left(\frac{\omega}{c}|\bx-\bx'|\right),\label{green.hank}
 \end{equation}
 where $\omega$ is temporal frequency and $H_0^{(1)}$ is the Hankel function of the first kind.
To reduce notational burden, we assume an implicit frequency dependence in all terms (unless otherwise specified). 
Thus the wavefield $\phi(\bx)$ excited by sources $S(\bx')$ is described by
 \begin{equation}
 \phi(\bx) = \int d\bx'\,G(\bx, \bx')\,S(\bx').\label{green.eq}
 \end{equation}
 Thus by treating $S(\bx')$ is a complex random variable, we may artificially generate temporal sequences
 of seismic noise with very little computational cost. Upon computing these displacement traces, we
investigate the properties of cross-correlations subject to generalized source distributions and signal processing methods.
The correlation in Fourier domain~(\ref{correl}) may be rewritten in terms of Green's functions
and sources
\begin{equation}
\tilde\cc_{\alpha\beta}(\omega) = \int d\bx'\int d\bx\,G^*(\bx_\alpha,\bx)\,G(\bx_\beta,\bx')\,S^*(\bx)\,S(\bx'),\label{exp.eq}
\end{equation}
If the variance of the complex random variable $S$ were to be bounded, the expectation
value of the cross-correlation~(\ref{expect}) exists and is given by
\begin{equation}
\cc_{\alpha\beta}(\omega) =\int d\bx'\int d\bx\,G^*(\bx_\alpha,\bx)\,G(\bx_\beta,\bx')\,\langle S^*(\bx)\,S(\bx')\rangle.\label{ccgen}
\end{equation}
For the demonstrations here, we assume that sources at all spatial points are described by independent and identically distributed random variables,
resulting in spatial decoherence, i.e.,
\begin{equation}
\langle S(\bx)\,S^*(\bx')\rangle = f(\bx)\,\delta(\bx-\bx').
\end{equation}

\begin{figure}
\centering
\includegraphics*[width=\linewidth]{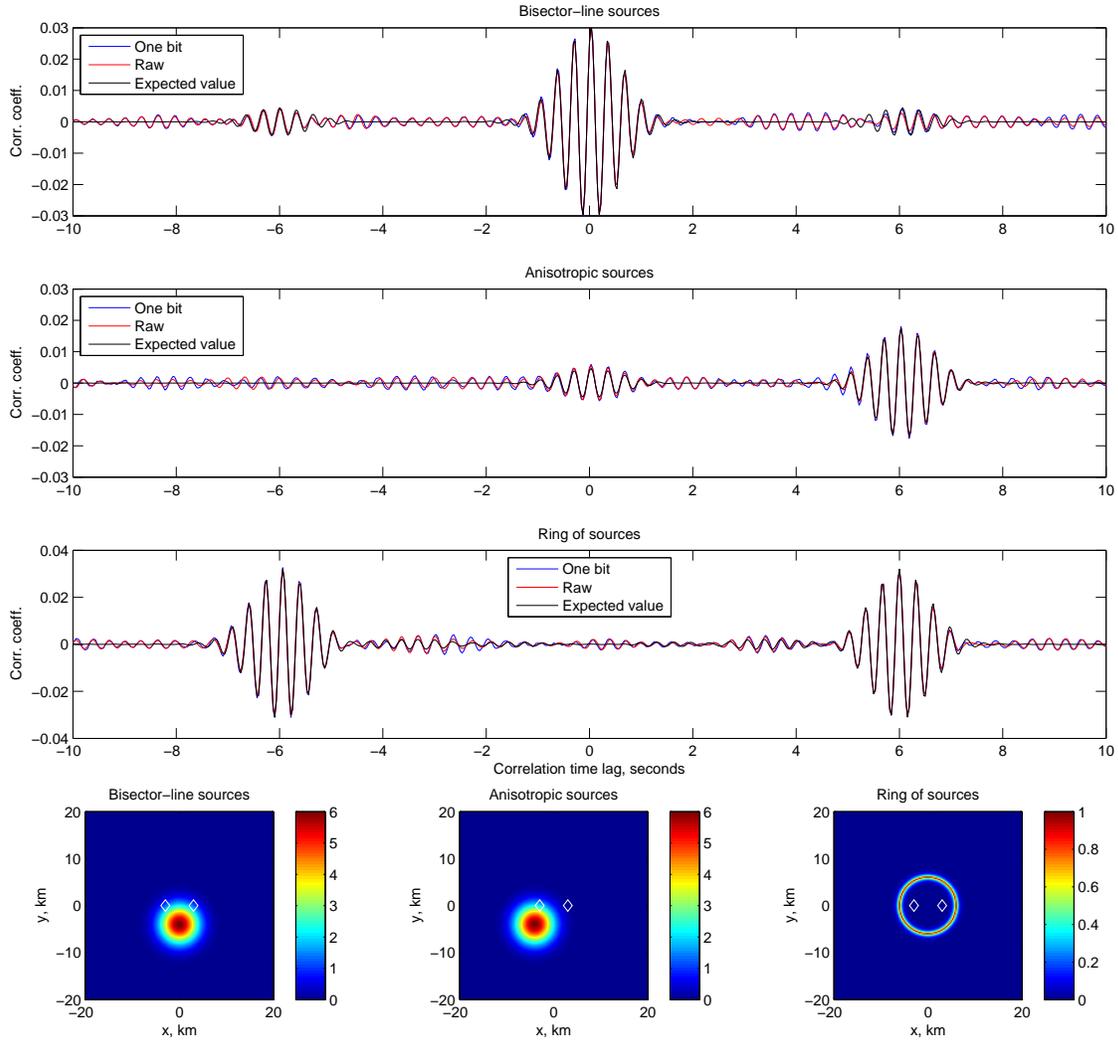}\vspace{-0.75cm}
\caption{Raw, one-bit and expected cross-correlation functions for configurations of source amplitudes (bottom). 
The correlation of the one-bit filtered variables is transformed according to equation~(\ref{onetoraw}) in order to allow for the comparison.
The source function
is populated with sequences of Gaussian-distributed complex random variables. Using equations~(\ref{green.eq}) and~(\ref{green.hank}), we 
generate seismic traces at the two stations and compute one-bit synthesized and raw cross-correlations. 12000 realizations of
this process are used in order to obtain this relatively high signal-to-noise result. The statistical amplitude configuration of sources is held stationary over all the realizations.
Since the source amplitudes obey Gaussian statistics, the expectation value of the cross-correlation exists and is computed
using equation~(\ref{ccgen}).  
\label{gaussian_so}}
\end{figure}

In order to generate realizations of seismic traces, we populate the source function with sequences of independent and identically distributed random variables $\{X\}, \{Y\}$ $S(\bx,\omega) = \{X\} + i\{Y\}$.
The seismic displacements at a given station is simply the product of Green's function between the source location and the realization $S(\bx,\omega)$. In this
manner, one can investigate the impact of different statistical distributions of sources on one-bit and raw cross-correlations. 
In Figure~\ref{gaussian_so}, we consider a pair of stations illuminated by two spatially non-uniform arrangements of sources, placed along the bisector line and
off-axis in two cases. We create 12000 realizations of sources and each realization of seismic traces are raw and one-bit correlated. These correlations are then
summed over all these realizations and the one-bit correlations are converted according to relation~(\ref{onetoraw}). The correlations in Figure~\ref{gaussian_so} show that
both one-bit and raw converge to the expectation value although they are both noisy.

  To further test the method, we consider sources that spatially non-stationary Gaussian, with and without large-amplitude perturbations (`earthquakes').
  We draw realizations of the non-Gaussian perturbations from the heavy-tailed L\'{e}vy alpha-stable distribution.
  
  \subsection{Spatially inhomogeneous and non-stationary Gaussian sources}
\begin{figure}
\centering
\includegraphics*[width=\linewidth]{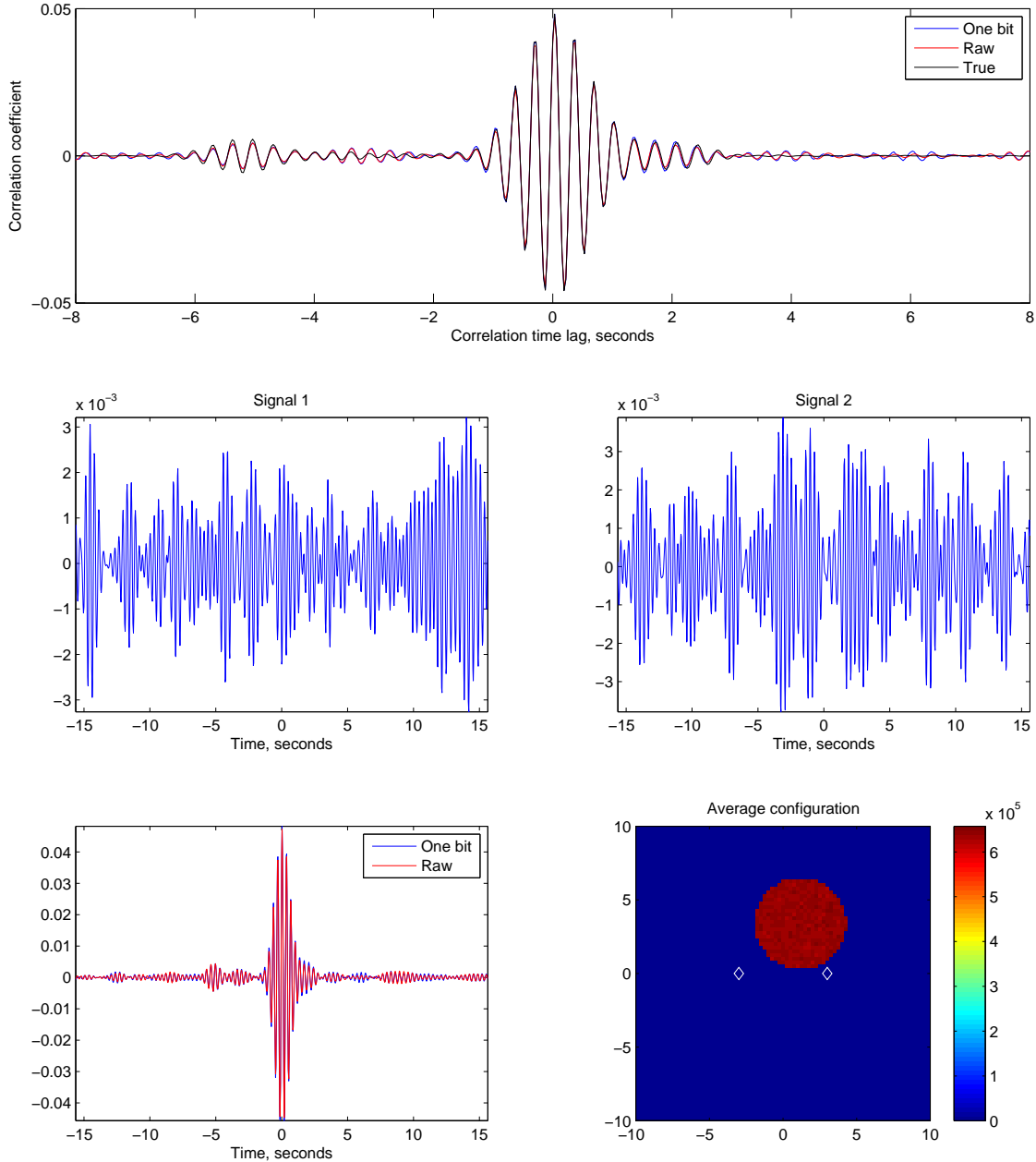}
\caption{In every realization, we choose a random set of source locations within the region of the disc (bottom right panel). Thus the sources are
spatially non-stationary, jumping from one set of points to another in each realization. The number of source points is chosen randomly and 
therefore varies from realization to the next.
  This choice is made to mimic a plausible scenario where sources both spatially and temporally stochastic.
We compute the traces as recorded by
the stations and compute one-bit and raw correlations using equation~(\ref{green.eq}). Sample raw traces drawn from one realization 
recorded at a pair of stations are shown  in the middle panels. These are considered representative of an average record.
In order to make the comparison, the one-bit correlation is converted according to equation~(\ref{onetoraw}).
We obtain an ``average configuration" by summing the spatial arrangement
of sources over all the realizations. The expectation value of the cross-correlation (assumed to exist) is then computed using this
average configuration and compared with the one-bit and raw correlations. It is seen that the three quantities are asymptotically 
identical and differences may attributed to measurement noise, i.e., partial convergence to the expectation value. 
\label{gaussian_nonstat}}
\end{figure}

  In this experiment, we perform 20000 realizations in which sources are assigned to a randomly chosen set of spatial locations. 
 Each realization consists of a randomly chosen number of source points within the disc shaped region shown on the bottom-right 
 panel of Figure~\ref{gaussian_nonstat}. Thus the sources `jump around' from one realization to the next.
  This is done to mimic a plausible scenario where sources both spatially and temporally stochastic.
  We assume as before that the sources are spatially incoherent, i.e., that they are spatially uncorrelated. 
  For each source, we populate its frequency domain representation with complex Gaussian random variables and using Green's theorem~(\ref{green.eq}), generate
  seismic traces at the two stations. We average the spatial arrangements of sources over all the realizations to obtain the
  ``average configuration" as shown in Figure~\ref{gaussian_nonstat}. The expectation value of the cross-correlation is then computed using equation~(\ref{ccgen})
  and compared with the raw and one-bit filtered correlations. In order to make the comparison, we apply the transfer
function~(\ref{onetoraw}). The agreement between the three curves is excellent.

\subsection{`Earthquake' perturbations}
\begin{figure}
\centering
\includegraphics*[width=\linewidth]{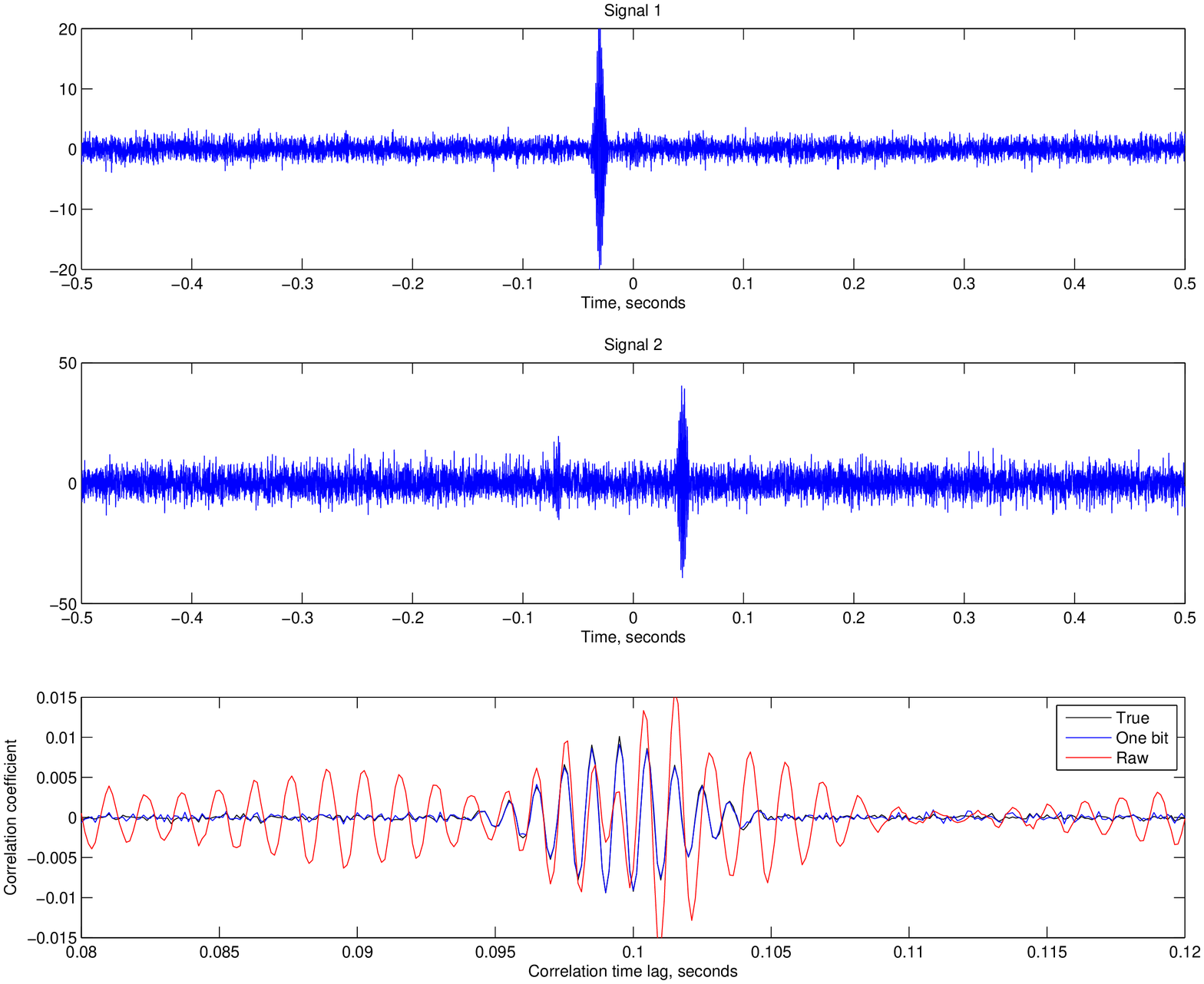}
\caption{The impact of earthquake-type sources on cross-correlations. We generate band-limited
bivariate Gaussian random variables, whose correlation function we term as `true'. Subsequently,
we add an `earthquake' by including a transient large-amplitude pulse in the signals (see the upper two panels). Raw and one-bit
correlations (converted using Eq.~[\ref{onetoraw}]) of these signals are computed. All in all, we compute and average these quantities over 2000 realizations.
It is seen that the one-bit correlation is nearly indistinguishable
from the ``true" whereas the raw is phase shifted and noisy. Amplitudes of the `earthquakes' are chosen from Cauchy distribution,
which belongs to the L\'{e}vy-alpha-stable family of distributions. Such statistics possess unbounded variances and consequently,
the expectation value of the raw cross-correlation is non-existent. The one-bit filter, which clips the tail of statistical distribution,
makes it possible to measure the noise correlation.
\label{deterministic}}
\end{figure}

In this experiment, we generate band-limited realizations of bivariate Gaussian random variables. These are likened to noise-source
generated seismic signals. We correlate these signals and term them the `true correlation'. Seismic records contain noise and 
earthquake related signals alike and consequently, in order to model the latter, we introduce intermittent large-amplitude transient oscillatory
signals into the measurements. In the upper two panels of Figure~\ref{deterministic}, we show an example of a stochastic 
realization with a superposed `earthquake'. The amplitude of the earthquake measurements is chosen randomly from a
Cauchy distribution, which belongs to the L\'{e}vy-alpha-stable family. Random variables from this family have
unbounded variances and consequently, correlation functions of such stochastic process are non-existent.
Raw- and one-bit correlations are computed and averaged over 8000 realizations, and we compare them
to the true correlation in the lowest panel of Figure~\ref{deterministic}. One-bit and true correlations are much more similar to each
other than the raw correlation, which is noisy and appears to be out of phase. Note that because the `earthquake' amplitudes are
Cauchy distributed, the raw correlation does not converge to an expectation value. In contrast, the one-bit correlation is well
behaved, with the one-bit filter clipping the tail of the distribution and large-amplitude events are weighted to the same extent
as noise. Consequently, the one-bit and true correlations are very similar.

\begin{figure}
\centering
\includegraphics*[width=\linewidth]{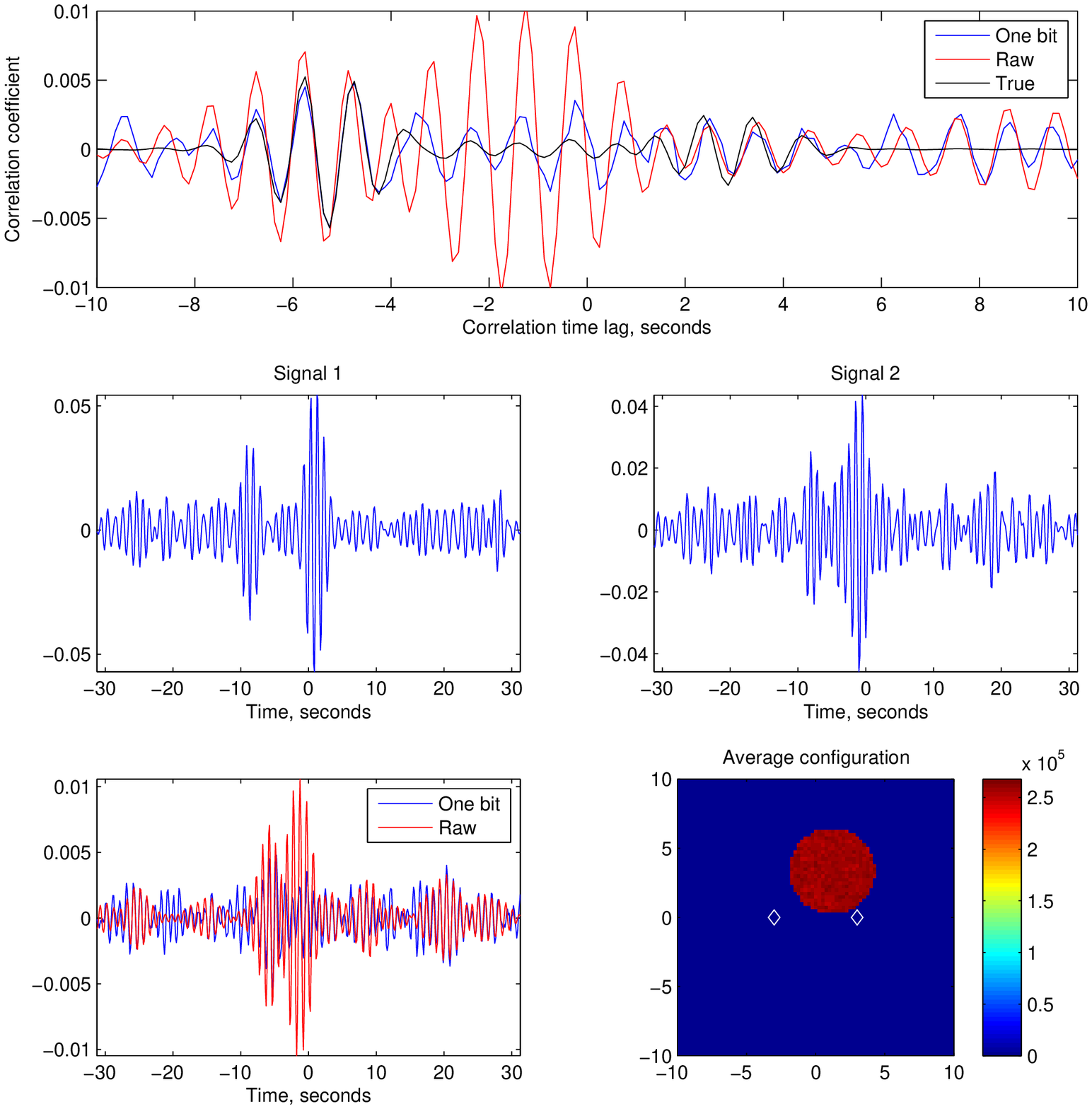}
\caption{The impact of earthquake-type sources on cross-correlations. 
In every realization, we choose a random set of source locations within the region of the disc. Thus the sources are
spatially non-stationary, jumping from one set of points to another in each realization.
We also add one `earthquake' per realization, which is effectively a large amplitude source (larger than the standard
deviation of the noise by a factor of at most 4), where the amplitude
is chosen from a Cauchy distribution.
We compute the traces as recorded by
the stations and compute one-bit and raw correlations using equation~(\ref{green.eq}); 
a realization of these traces is shown in the middle panels. 
The `true' correlation is computed using the `earthquake'-free seismic traces and
the one-bit correlation is converted
using equation~(\ref{onetoraw}) in order to make this comparison. Much as in Figure~\ref{deterministic}, it is seen
that the one-bit and `true' correlations are much closer to being in agreement whereas the raw correlation is out of phase.
As before, the expectation value of the raw correlation is non-existent because of the unbounded variance
of random variables belonging to the Cauchy distribution.
\label{deterministic2}}
\end{figure}

\section{Conclusions}
We have studied and categorized the statistical properties of one-bit and raw correlations. 
Overall, our assessment is that the one-bit
correlation is a more stable measurement, able to handle large-amplitude signals due to sporadic sources, with
signal-to-noise characteristics superior in the presence of these perturbations and at worst, comparable to the raw correlation.
A summary of the assumptions and properties of the method are listed
\begin{itemize}
\item The theory applies to both stationary and non-stationary random processes, which obey either Gaussian or a broad range of other distributions (see appendix~\ref{corrgen}),
\item If these assumptions are true, correlations obtained through one-bit processing (with the transfer function) accurately recover the correlation function,
\item Simple tests with non-stationary processes show that the `true' correlation (associated with small-amplitude fluctuations) is well recovered,
\item The short-window spectral-whitening method \citep[a variant of the method discussed by][]{seats12} converges at a rate similar to that of one-bit correlations for
the cases we consider here,
\item Information is lost when one-bit filtering is applied and the signal-to-noise ratio will consequently be lower than when computing correlations using raw data
for purely Gaussian distributed fluctuations. However, when the raw data are punctuated by large-amplitude transient events, the one-bit filtered
correlations converge to the correct expectation value,
\item The one-bit correlations can be directly assimilated and interpreted in the framework of the adjoint method, as discussed by, e.g., \cite{tromp10, hanasoge12_sources}.
\end{itemize}
Although a more comprehensive study is needed to fully characterize the signal-to-noise losses and gains, our tests show 
that using the one-bit filter to process terrestrial seismic noise can be very beneficial. 
One may recover the true noise correlation by first estimating the one-bit correlation, applying transfer function~(\ref{onetoraw}),
and multiplying the resultant by the standard deviations of the measurements at the two stations. This way, one can reconstruct
the correct amplitudes of the cross correlations and this information is not lost.
The elegance of using one-bit filtering lies in being able to effortlessly move between the filtered and real domains.
This implies that theoretical developments in aid of interpreting noise correlation measurements (e.g., \cite{tromp10, hanasoge12_sources}) need no altering,
and the data processing techniques that are currently widely in use are fully justifiable. We conclude with an algorithm to leverage these benefits:
\begin{itemize}
\item Measure $\sigma_i$, of the noise records as best as possible (i.e., by ignoring large-amplitude spikes)
\item Convert real measurements at pairs of stations $i,j$ into digital signals (apply the one-bit filter)
\item Correlate these signals, and determine the one-bit correlation function $\tilde\rho^1_{ij}(\tau)$
\item Use transfer function~\ref{onetoraw} to move between the filtered domain to the real domain (i.e., ${\tilde\rho}_{ij}(\tau) = \sin(\pi\tilde\rho^1_{ij}/2)$)
\item Renormalize ${\tilde\rho}_{ij}(\tau)$ using the product of the measured standard deviations $\sigma_i\sigma_j$ to recover the cross-correlation, i.e.
$\tilde\cc_{ij}(\tau) = \sigma_i\sigma_j{\tilde\rho}_{ij}(\tau)$
\item These measurements can now directly be assimilated into the inversions according to the theory of e.g., \cite{tromp10, hanasoge12_sources}.
\end{itemize}

An important issue is whether the method of one-bit filtering remains effective when the distribution of seismic noise fluctuations is non-Gaussian.
Does equation~(\ref{onetoraw}), which allows us to move so easily between the filtered and real domains, continue to remain valid? 
In appendix~\ref{corrgen}, we have discussed more generalized distributions (i.e., non-Gaussian fluctuations) and address a range of non-linear modulators. For the cases we have studied,
where the fluctuations possessed exponential/ other heavy-tailed distributions, we find that relation~(\ref{onetoraw}) still remains very effective. 

\section*{Acknowledgements}
S. M. H. is funded by NASA grant NNX11AB63G and thanks Courant Institute, New York University 
for its hospitality and G\"{o}ran Ekstr\"{o}m for useful conversations. 
We also warmly thank Cornelis Weemstra for his careful reading and considerable help in improving the manuscript
and an anonymous referee for helping us tighten our narrative. 
This work is an effort to understand cross-correlations 
in helioseismology in the context of DFG CRC 963  ÒAstrophysical Flow Instabilities and TurbulenceÓ. 

\bibliographystyle{gji}

\appendix

\section{Fourier Convention}\label{convention}
We apply the Fourier convention below
\begin{eqnarray}
\int_{-\infty}^\infty dt~e^{i\omega t}~ g(t) &=& {\hat g}(\omega) ,\\
\int_{-\infty}^\infty dt~e^{i\omega t} &=& 2\pi~\delta(\omega),\label{inv.fourier}\\
\frac{1}{2\pi}\int_{-\infty}^{\infty} d\omega~e^{-i\omega t}~ {\hat g}(\omega) &=& g(t),\\
\int_{-\infty}^\infty d\omega~e^{-i\omega t} &=&  2\pi~\delta(t),
\end{eqnarray}
where $g(t), {\hat g}(\omega)$ are a Fourier transform pair. 
Cross-correlations and convolutions in the Fourier and temporal domain are described by
\begin{equation}
h(t) = \int_{-\infty}^{\infty} dt'~ f(t')~ g(t+t') \Longleftrightarrow {\hat h}(\omega) = {\hat f}^*(\omega)~{\hat g}(\omega),\label{cross.c} \\
\end{equation}
\begin{equation}
h(t) = \int_{-\infty}^{\infty} dt'~ f(t')~ g(t-t') \Longleftrightarrow {\hat h}(\omega) =  {\hat f}(\omega)~{\hat g}(\omega).
\end{equation}
For real functions $f(t), g(t)$, we also have
\begin{equation}
\int_{-\infty}^\infty dt~f(t)~g(t) = \frac{1}{2\pi}\int_{-\infty}^{\infty} d\omega~{\hat f}^*(\omega)~{\hat g}(\omega) = \frac{1}{2\pi}\int_{-\infty}^{\infty} d\omega~{\hat f}(\omega)~{\hat g}^*(\omega).
\end{equation}

\section{General framework for assessing the statistical properties of nonlinear modulators}\label{corrgen}
The goal here is to obtain general formulas linking the the cross-correlation function
\[\CC_x(t_1,t_2)=\langle x_1(t_1) x_2(t_2)\rangle ,\] 
for two signals, $x_1, x_2$ measured at different stations,  to the cross-correlation  function of modulated signals 
\[\CC_H(x)(t_1,t_2)=\langle H(x_1(t_1))H( x_2(t_2))\rangle ,\] 
$H$ is a nonlinear modulator with known characteristics.
We seek an exact or approximate link allowing to recover $\CC_s$ from $\CC_{H(x)}$. This framework is much more general than the one outlined in the main text due to the following reasons:
\begin{itemize}
\vspace{-.2cm}\item No stationarity in the measured signals, $x_1$ and $x_2$ is assumed for the most part,
\item The measured signals, $x_1$ and $x_2$ are assumed to contain both the seismic noise component $s_i$ and some unwanted sources of tectonic origin, i.e., $x_i = s_i+\r_i$; the assumption here that the processes $\r_i$ have non-Gaussian statistics with fat tails indicative of intermittent events, e.g., events of tectonic origin.
\item A wide class of nonlinear modulators $H$ is discussed; the one-bit digitizer discussed in the main text is but one member of the class.
\end{itemize}

We seek an exact or approximate link allowing to recover the information about the seismic noise, $\CC_s$, from the modulated correlation function $\CC_{H(x)}$ of the modulated signal which is corrupted by the tectonic sources.
   Motivated by the earlier work in \cite{rice44,midd48,hall}, we express the correlation function $\CC_{H(x)}$ 
of the modulated signal via an integral involving the characteristics of the modulator and the joint characteristic function of the input signals. 
Exact analytical or approximate formulas for $\CC_{H(x)}$ can be derived if the joint characteristic function of the input signals can be represented as a series of appropriately factorized terms; such a factorized series representation is possible for a large class of Gaussian and non-Gaussian processes, as shown below. In particular, in \S\ref{onetoraw2} we provide an alternative derivation of the simple trigonometric link (\ref{onetoraw}) between the Gaussian seismic noise correlations and its modulated correlations. 
We then show that for a large class of bivariate non-Gaussian processes the correlation function $\CC_{H(x)}{=}\langle H\big(x_1(t)\big)H\big(x_2(t)\big)\rangle $ of the filtered corrupted signals  is insensitive to the correlations between the corrupting processes $\r_i(t)$ and it is the same as for uncorrelated processes $\r_i(t)$ with $\langle\r_1(t_1)\r_2(t_2)\rangle=0$. Most importantly, the relationship between the correlation function of the modulated signal $\CC_{H(x)}$ and the correlation function of the Gaussian seismic noise $\CC_s$ is given, to a very good approximation, by the  simple trigonometric formula (\ref{onetoraw}) established  in the Gaussian context.

\subsection{General formula for the cross-correlation function}\label{gen_form}
Here, we consider the statistics of the modulated  signal $H(x(t))$ given the known characteristics of the input signal $x(t)$ and the characteristics of the modulator $H$ which we assume to have the Laplace transform representation 
\begin{align}\label{HL}
H(x) = \frac{1}{2\pi \i}\int_{C} \hat{H}(z)e^{zx}\rd z,
\end{align}
where $\hat H(z)=\int_{0}^{\infty} H(x)e^{-zx}\rd x$ and $C$ is a suitable contour in the complex plane.
The main focus is on nonlinear modulators and it is well known that a large class of such devices  can be represented in such a way (e.g., \cite{rice44,midd48}).
The key feature of the integral representation (\ref{HL}) is that the cross-correlation function of the modulated signals $H(x_1)$ and $H(x_2)$ can be expressed via an integral involving the joint characteristic function of the input  as 
\begin{align}\label{corr}
\CC_{H(x)}(t_1, t_2) &\equiv \big\langle H(x_1)H(x_2)\big\rangle = \int_{-\infty}^{\infty}\int_{-\infty}^{\infty}H(x_1)H(x_2)p(x_1,x_2)\rd x_1\rd x_2 \notag\\
&=\frac{1}{(2\pi \i)^2 } \int_C \hat H(z_1)\rd z_1\int_{C}\hat H(z_2)\rd z_2 \int_{-\infty}^{\infty}\int_{-\infty}^{\infty} p(x_1,x_2)e^{z_1x_1+z_2x_2}\,\rd x_1\rd x_2,
\end{align} 
where $p(x_1,x_2)$ is the joint  density associated with the input pair  $\big(x_1(t_1), x_2(t_2)\big)$; here and below, we skip the explicit dependence on $t_1,t_2$ in the joint density $p(x_1,x_2,t_1,t_2)$ in order to simplify the notation.
Note that the last integral in (\ref{corr}) is the joint characteristic function of $p(x_1,x_2)$ given by 
\begin{equation}\label{char}
\chi(z_1,z_2) =  \textrm{E}_{x_1,x_2}\big[e^{z_1x_1+z_2x_2}\big]=\int_{-\infty}^{\infty}\int_{-\infty}^{\infty} p(x_1,x_2)e^{z_1x_1+z_2x_2}\,\rd x_1\rd x_2,
\end{equation}
so that we can express the cross-correlation of the modulated  signal (\ref{corr}) through the Laplace transform of the filter and the joint characteristic function of the input signal  as
\begin{align}\label{C}
\CC_{H(x)}(t_1, t_2) 
&=\frac{1}{(2\pi \i)^2 } \int_C \hat H(z_1)\rd z_1\int_{C}\hat H(z_2) \,\chi(z_1,z_2)\,\rd z_2.
\end{align}

The attractive property of the exact formula in (\ref{C}) is that it splits the properties of the modulator, represented by the terms involving $\hat H$, from the properties of the input signal encoded in the characteristic function $\chi$. 

\medskip
Here, we assume that the input signals $x_1$ and $x_2$ contain the signal of interest and perturbations; our goal is to examine the ability of the modulator $H$  to recover the statistical properties of the signal $s(t)$ from the input $x(t)$ which is corrupted  by some unwanted processes  $\r(t)$; thus, we write the input signal as 
\begin{equation}\label{sn}
x_i(t) = s_i(t)+\r_i(t), \qquad\quad   E\big[s_i(t)\r_j(t)\big]=0, \quad \textrm{for} \;\; i,j\in \{1,2\},
\end{equation}
where the perturbations $\r(t)$ are independent of $s(t)$.  
Given the form of the input as in (\ref{sn}), its joint characteristic function factorizes as 
\begin{align}\label{chix}
\chi_x(z_1,z_2)&=\chi_s(z_1,z_2)\chi_{\r}(z_1,z_2).
\end{align} 

\subsection{Special cases of the general framework}
The utility of the general framework described above becomes apparent when the joint characteristic functions of the truth signal and of the corrupting process can be factorized as 
\begin{align}\label{fact}
\chi_s(z_1,t_1,z_2,t_2) = \sum_{k_s} a_{k_s}(z_1,t_1)b_{k_s}(z_2,t_2), \qquad \chi_{\r}(z_1,t_1,z_2,t_2) = \sum_k c_{k}(z_1,t_1)d_{k}(z_2,t_2),
\end{align}
so that 
\begin{align}\label{chi_fact}
&\chi(z_1,t_1,z_2,t_2) = \chi_s(z_1,t_1,z_2,t_2) \chi_{\r}(z_1,t_1,z_2,t_2) \notag\\[.3cm]
&\hspace{1cm}= \sum_{k_s}\sum_{k_{r}}\Big(a_{k_s}(z_1,t_1)c_{k_r}(z_1,t_1)\Big) \,\Big(b_{k_s}(z_2,t_2)d_{k_r}(z_2,t_2)\Big).
\end{align}

\medskip
Given the factorization in (\ref{chi_fact}), the autocorrelation of the modulated signal (\ref{C}) becomes 
\begin{align}\label{CGH}
\CC_{H(x)}(t_1,t_2)=\sum_{k_s}\sum_{k_{r}}\textfrak{G}_{k_s, k_{r}}(t_1)\textfrak{H}_{k_s, k_{r}}(t_2)
\end{align}
where
\begin{align}
\textfrak{G}_{k_s, k_{r}}(t_1)=\frac{1}{2\pi \i}\int_C \hat H(z_1)a_{k_s}(z_1,t_1)c_{k_r}(z_1,t_1)\rd z_1,\\[.3cm]
\textfrak{H}_{k_s, k_{r}}(t_2)=\frac{1}{2\pi \i}\int_C \hat H(z_2)b_{k_s}(z_2,t_2)d_{k_r}(z_2,t_2)\rd z_2.
\end{align}
The form of $\textfrak{G}$ and $\textfrak{H}$ in (\ref{CGH}) depends on the modulator $H$ and the statistics of the input signal. A particular class of modulators commonly used in signal processing analyzed is introduced below.

\subsubsection{The family of nonlinear modulators}\label{modul}

We consider a one parameter class of nonlinear modulators described by 
\begin{align}\label{H}
H_{\alpha}(x) = \begin{cases} \hspace{.5cm}x^\alpha &\quad\textrm{for} \quad x>0,\\
  \hspace{.5cm}0 & \quad  \textrm{for}\quad  x=0,\\ 
  -(-x)^\alpha &\quad \textrm{for}\quad x<0,\end{cases}
\end{align}
where the constant $\alpha\geqslant 0$ parameterizes the  family. 
The filter $H_{\alpha}$ in (\ref{H}) has the following integral representation (cf. (\ref{HL}))
\begin{align}\label{Hpm}
H_{\alpha}(x) = \frac{1}{2\pi \i}\int_{C} \hat{H}_{\alpha}(z)e^{zx}\rd z=\frac{1}{2\pi \i}\int_{C_+} \!\hat{H}_{\alpha}^+(z)e^{zx}\rd z+\frac{1}{2\pi \i}\int_{C_-} \!\hat{H}_{\alpha}^-(z)e^{zx}\rd z,
\end{align}
where we exploit the Laplace transform of $H$ with the contours $C_+,\;C_-$ given, respectively, by 
\begin{equation}
C_+ = \epsilon +\i v\qquad C_- = -\epsilon+\i v,\qquad  v\in\RR,
\end{equation}
and the transforms, $\hat H_{\alpha}^+, \;\hat H_{\alpha}^-$, are expressed through the Gamma function, $\Gamma(z) = \int_0^\infty e^{-t}t^{z-1}\rd t$, as follows:
\begin{equation}\label{hatH}
\hat H_{\alpha}^+(z) = \frac{\Gamma(\alpha+1)}{z^{\alpha+1}}, \qquad \hat H_{\alpha}^-(z) = -\frac{\Gamma(\alpha+1)}{(-z)^{\alpha+1}}.
\end{equation}
The two-parameter family of modulators in (\ref{H}) contains a large class of modulators ranging from a trivial linear case $H_{1}$ to the one-bit modulator $H_{0}$ the properties of which are the main focus here.

\subsection{Cross-correlation functions for one-bit filtered Gaussian processes}\label{onetoraw2}
 Here, we show that in the particular case of uncorrupted Gaussian signal, $x_i(t)=s_i(t), \;i=1,2$, modulated by the so-called ideal limiter~$H_{0}$, the correlation functions of the input $\CC_s$ and of the modulated output $\CC_{H(s)}$ are related by 
 \begin{align}\label{rhrs}
\boxed{\CC_{H(s)}(\tau) =\frac{2}{\pi} \arcsin\left(\frac{\CC_s(\tau)}{\sigma_{1}\sigma_{2}}\right),}
\end{align} 
where $\sigma_{i}^2= \langle s_i^2\rangle-\langle s_i\rangle^2$. This result obtained by reducing the general non-Gaussian framework of \S\ref{gen_form}-\ref{modul} is complementary to the derivation discussed in section~\ref{gauss.one} which requires the Gaussianity assumptions from the outset. 
Consider a `canonical' Gaussian process with autocorrelation $\CC_s$ and the joint characteristic function given by 
 \begin{equation}\label{chig}
\chi_s(z_1,z_2,t_1,t_2) =  \exp\left\{\frac{1}{2}\Big[\sigma^2_1z_1^2+\sigma^2_2z_2^2+2\,\CC_s(t_1,t_2)z_1z_2\Big]\right\}=\sum_{k=0}^\infty \frac{\CC_s^k(t_1,t_2)}{k!} e^{\frac{1}{2}\sigma_{1}^2z_1^2}z_1^k \,e^{\frac{1}{2}\sigma_{2}^2z_2^2}z^k_2,
\end{equation}
where the second equality is due to the expansion
\begin{equation}
\exp\big[\CC(t_1,t_2)z_1z_2\big] = \sum_{k=0}^\infty \frac{\CC^k(t_1,t_2)}{k!}z_1^kz_2^k.
\end{equation}
Consequently, the correlation function (\ref{CGH}) of the digitized signal is given by 
\begin{align}\label{chgauss}
\CC_{H(s)}(t_1, t_2) =\sum_{k=0}^{\infty}\mathscr{G}^{\alpha}_{k}(t_1)\mathscr{G}^{\alpha}_{k}(t_2)\frac{C_{s}^k(t_1,t_2)}{k!} \,,
\end{align}
where
\begin{align}
\mathscr{G}^{\alpha}_{k}(t)&=\frac{1}{2\pi \i}\int_C \hat H_{\alpha}(z)e^{\frac{1}{2}z^2\sigma(t)^2}\!z^{k}\,\rd z\notag\\
&=\frac{1}{2\pi \i}\int_{C^+} e^{\frac{1}{2}z^2\sigma(t)^2}\!z^{k-1}\,\rd z-\frac{1}{2\pi \i}\int_{C^-} e^{\frac{1}{2}z^2\sigma(t)^2}\!z^{k-1}\,\rd z\label{gauss_cnt},
\end{align}

The contour integrals in (\ref{gauss_cnt}) above can be evaluated in a standard fashion noticing (we skip details here) that in such a case both contours can be shifted to the imaginary line $(-\i\infty,\; \i \infty)$ so that upon substituting $z=\i w$ in (\ref{gauss_cnt}), we obtain 
\begin{align}
\mathscr{G}_{k}= \begin{cases} \displaystyle \frac{2^{k/2}}{\sigma_1^{k}}\,\frac{1}{\Gamma\left(1-k/2\right)}\quad &\textrm{for} \quad k \quad \textrm{odd}, \\[.3cm] \hspace{1cm}0 \quad &\textrm{for} \quad k \quad \textrm{even},\end{cases}
\end{align}
and the autocorrelation function (\ref{chgauss}) becomes 
\begin{align}\label{chs}
\CC_{H(s)}(t_1, t_2) &=\sum_{k=0}^{\infty}\mathscr{G}_{k}\mathscr{H}_{k}\,C_{k}(t_1,t_2)=\sum_{n=0}^{\infty} \frac{2^{2n+1}}{(2n+1)!\,\Gamma^2\left(\frac{1}{2}-n\right)}\frac{\CC_s^{2n+1}(t_1,t_2)}{\sigma_1^{2n+1}\sigma_2^{2n+1}}.
\end{align}
Further simplifications in (\ref{chs}) can be achieved  using the fact that 
\[\Gamma\left(\textstyle\frac{1}{2}-n\right) = \frac{\sqrt{\pi}(-1)^22^{2n}n!}{(2n)!},\]
for any integer $n$ which leads to 
\begin{align}\label{CC_gss_nonst}
\CC_{H(s)}(t_1, t_2) 
&=\frac{2}{\pi}\sum_{n=0}^{\infty} \frac{(2n)!}{2^{2n}(2n+1)(n!)^2}\frac{\CC_s^k(t_1,t_2)}{\sigma_1^{k}\sigma_2^k}=\frac{2}{\pi} \arcsin\left(\frac{\CC_s(t_1,t_2)}{\sigma_1\sigma_2}\right).
\end{align}

\noindent If the signals $s_1,s_2$ are stationary, i.e., $\CC_s(t,t+\tau)=\CC_s(0,\tau)$ for all $t$, the above expression simplifies to 
\begin{align}
\CC_{H(s)}(\tau) =\frac{2}{\pi} \arcsin\left(\frac{\CC_s(\tau)}{\sigma_1\sigma_2}\right),
\end{align}
as claimed at the beginning of this section.

\subsection{Cross-correlation functions for a signal consisting of Gaussian and non-Gaussian components}\label{gss_ngss}
Here, we consider non-Gaussian perturbations to the seismic noise signals $s_1,s_2$ measured at two different stations. We show that for a large class of densities the correlation function $\CC_{H(x)}{=}\langle H\big(x_1(t)\big)H\big(x_2(t)\big)\rangle $ of the filtered corrupted signals  is insensitive to the correlations between the corrupting processes $\r_i(t)$ and the same as for uncorrelated processes $\r_i(t)$ with $\langle\r_1(t_1)\r_2(t_2)\rangle=0$. Most importantly the relationship between the correlation function of the filtered corrupted signal $\CC_{H(x)}$ and the correlation function of the Gaussian seismic noise $\CC_s$ is given, to a very good approximation, by the simple trigonometric formula (\ref{rhrs}).

Consider the two input signals, $x_i(t)=s_i(t)+\r_i(t), i=1,2,$ to consist of the  Gaussian components $s_i(t)$  corrupted by processes $\r_i(t)$ with Cauchy-like density (at any fixed time)
\begin{equation}\label{dens_r}
f(\r) = \frac{2}{\pi}\frac{\gamma^3}{(\gamma^2+\r^2)^2}, \quad \gamma>0.
\end{equation} 
It is well known (e.g., \cite{gumbel60}) that construction of joint probability density of a bivariate process  with given marginal densities, which is necessary for deriving the characteristic function in (\ref{char}), is non-unique. One possible way of constructing such a joint density with arbitrary correlation function is via  the following formula
\begin{equation}\label{p_gen1}
p(x,y)=f(x)f(y)+\sum_i g_i(x)g_i(y),
\end{equation}
where $g_i$ are odd functions and chosen such that $p(x,y)\geqslant 0$. Below we show that for this class of densities the correlation function of the corrupted signals $x_1(t), x_2(t)$ is insensitive to the correlations between the corrupting processes $\r_i(t)$ and the same as for the delta-correlated process $\r(t)$, i.e., $\textrm{E}[\r_1(t_1)\r_2(t_2)]=0$.

Consider, in particular, the following two-point density with the structure as in (\ref{p_gen1})
\begin{equation}\label{p_fF}
p(\r_1,\r_2)=f(\r_1)f(\r_2)\Big[1+\CC_\r(t_1,t_2)\tilde F^{-2}(2F(\r_1)-1)(2F(\r_2)-1)\Big],
\end{equation}
where $\CC_\r(t_1,t_2)$ is the cross-correlation of $\r_1$ and $\r_2$ and the marginal cumulative distribution function $F(x)$ is given by 
\begin{equation}
F(x)=\int_{-\infty}^x f(x')\rd x' = \frac{\gamma}{\pi}\frac{x}{\gamma^2+x^2}+\frac{1}{\pi}\textrm{arctg}\left(\frac{x}{\gamma}\right)+\frac{1}{2},
\end{equation}
and the normalization constant in (\ref{p_fF}) is $\tilde F = \int_{-\infty}^\infty xf(x)(2F(x)-1)\rd x=\frac{3\gamma}{2\pi}$.

The joint characteristic function for the Gaussian signal has the same form as in (\ref{chig}) and  
the joint characteristic  function of the perturbing processes $\r_1(t_1), \r_2(t_2)$ is given in the factorized form as 
 \begin{equation}\label{gchin}
\chi_{\r}(z_1,z_2) =  (1+\gamma_1|z_1|)(1+\gamma_2|z_2|)\exp\left\{-\gamma_1|z_1|+\gamma_2|z_2|\right\}+\CC_\r(t_1,t_2)\mathcal{F}(z_1)\mathcal{F}(z_2);
\end{equation}
the Fourier transforms $\mathcal{F}$ in (\ref{gchin}) have no analytical expressions and are computed numerically below. 
Consequently, the correlation function of the modulated input signal $x(t)$ signal is given by 
\begin{align}
&\CC_{H(x)}(t_1, t_2) =\sum_{k=0}^{\infty}\frac{\mathscr{G}^{(\alpha)}_{k}(\sigma_1,\gamma_1)\mathscr{G}^{(\alpha)}_{k}(\sigma_2,\gamma_2)}{k!}\,\CC_{s}^k(t_1,t_2) \\[.3cm]
&\hspace{3cm}+\sum_{k=0}^{\infty}\frac{\mathscr{H}^{(\alpha)}_{k}(\sigma_1,\gamma_1)\mathscr{H}^{(\alpha)}_{k}(\sigma_2,\gamma_2)}{k!}\,\CC_{s}^k(t_1,t_2)\CC_\r(t_1,t_2),
\end{align}
where
\begin{align}
\mathscr{G}^{(\alpha)}_{k}(\sigma,\gamma)&=\frac{1}{2\pi \i}\int_C \hat H_{\alpha}(z)e^{\frac{1}{2}z^2\sigma^2-\gamma|z|}(1+\gamma|z|)z^{k}\rd z,\label{GG}\\[.2cm]
\mathscr{H}^{(\alpha,h)}_{k}(\sigma,\gamma)&=\frac{1}{2\pi \i}\int_{C}\hat H_{\alpha}(z) \mathcal{F}(z)e^{\frac{1}{2}z^2\sigma^2}z^{k}\,\rd z.\label{HH}
\end{align}

\begin{figure}[t]
\centering
\includegraphics[width=8cm]{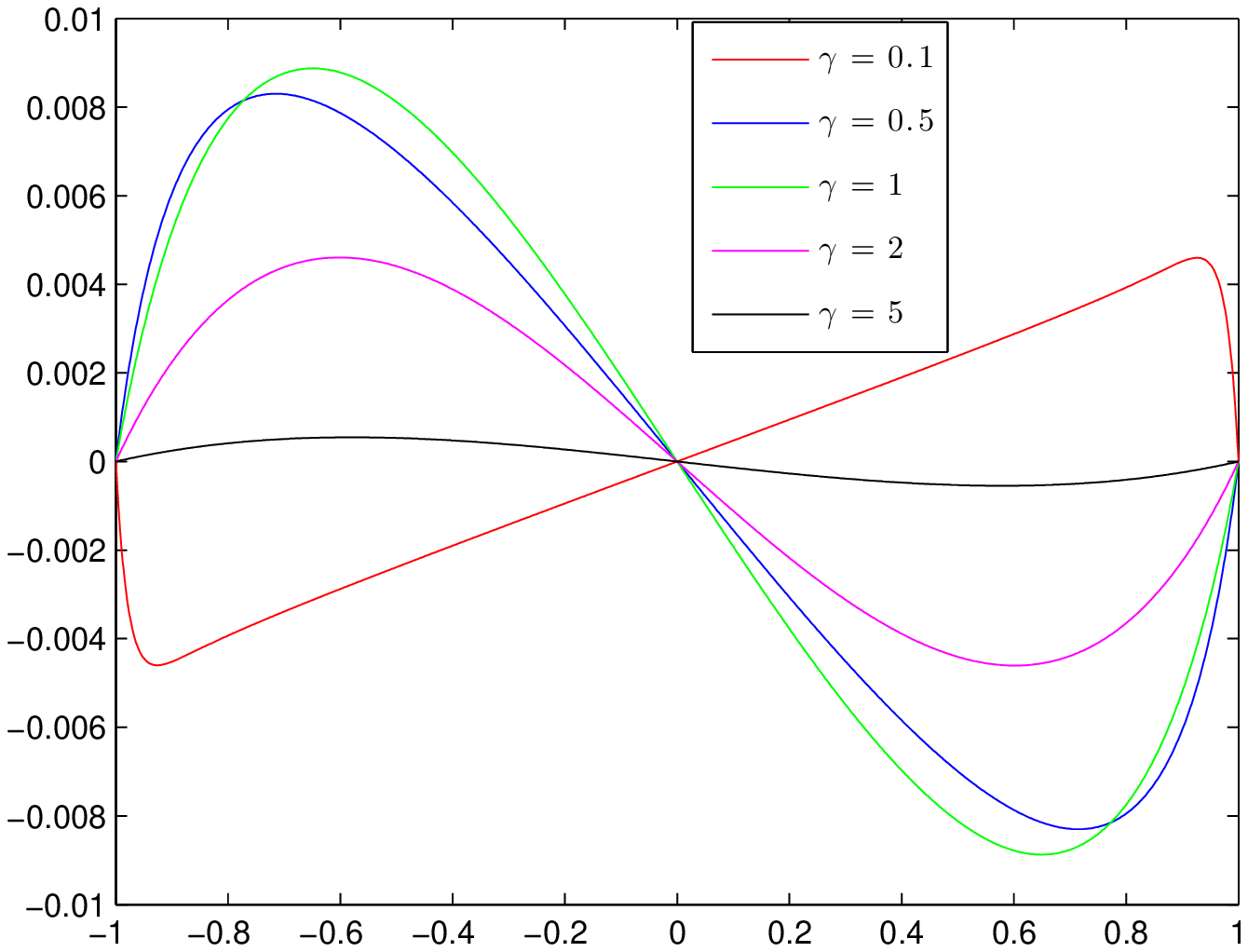}
\caption{Error in the approximation (\ref{chgauss5}) of the formula (\ref{chgauss3}); different curves show the residual $\CC_s(\tau)/\CC_s(0)-\sin\Big(\textstyle\frac{\pi}{2}\CC_{H(x)}(\tau)\Big)/\sin(\frac{\pi}{2} \CC_{H(x)}(0))$ for different values of the variance $\gamma^2$ of the non-Gaussian process $\r(t)$.}\label{error}
\end{figure}

\begin{figure}
\centering
\includegraphics[width=14cm]{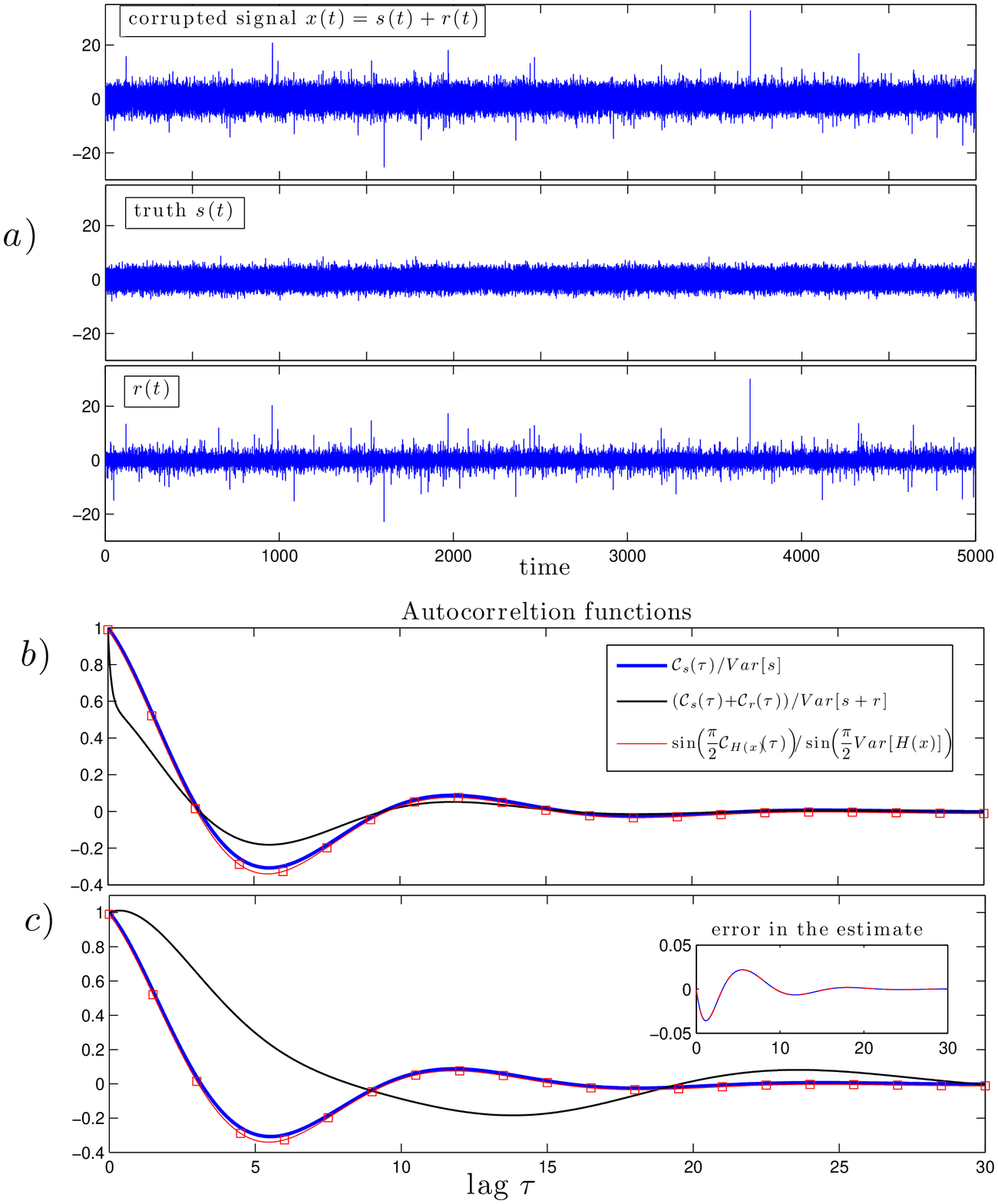}
\caption{Recovering the autocorrelation function of the signal $s(t)$ from the input $x(t){=}s(t){+}\r(t)$ corrupted by non-Gaussian process $\r(t)$ via the one-bit modulator $H_0$ (\ref{H}). a) Example of the corrupted Gaussian signal x(t) and its components $s(t)$ and $\r(t)$. b,c) Examples of normalized autocorrelation functions of the truth (blue), the corrupted input (black), and the recovered autocorrelation (red) of the truth via $\sin(\frac{\pi}{2} C_{H(x)})$ in (\ref{chgauss2}).  The inset b) shows the case when the non-Gaussian process with density (\ref{dens_r}) is white in time (i.e., $\CC_\r(\tau)=(1.5)^2\delta(\tau)$) and the correlation function of the truth is $\CC_s(\tau) \propto \exp(-0.2\tau)\cos(0.5t)$. The example in c) is for the same truth signal as in b) but the non-Gaussian process has a long correlation time with $\CC_\r(\tau) =\exp(-0.1\tau)\sin(0.3 t+0.5)$.}\label{exampl}
\end{figure}

The contour integrals in (\ref{GG})-(\ref{HH}) above can be evaluated in a standard fashion using (\ref{Hpm}) and (\ref{hatH}), and noticing (we skip details here) that in such a case both contours can be shifted to the imaginary line $(-\i\infty,\; \i \infty)$; thus, upon substituting (\ref{hatH}) for $\hat H_\alpha$ and $z=\i w$ in (\ref{GG}), (\ref{HH}), we obtain 
\begin{align}\label{ggauss1}
\mathscr{G}^{(\alpha)}_{k}(\sigma,\gamma)= \begin{cases} \displaystyle \frac{2\Gamma(\alpha+1)}{\pi \,k!}\sin \Big(\frac{\pi}{2}(k-\alpha)\Big)\int_{0}^\infty e^{-\frac{1}{2}\sigma^2w^2-\gamma w}(1+\gamma w)w^{k-\alpha-1}\rd w\quad &\textrm{for} \quad k \quad \textrm{odd}, \\[.5cm] \hspace{3cm}0 \quad &\textrm{for} \quad k \quad \textrm{even}.\end{cases}
\end{align}
and 
\begin{align}\label{ggauss2}
\mathscr{H}^{(\alpha)}_{k}(\sigma,\gamma)\propto\sin \Big(\frac{\pi}{2}(k-\alpha)\Big)\int_{0}^\infty \mathcal{F}(w)e^{-\frac{1}{2}\sigma^2w^2}w^{k-\alpha-1}\rd w=  0.\end{align}
The result in (\ref{ggauss2}) is due to the fact that $\mathcal{F}(w)$ is an odd function. Consequently, the correlation function in (\ref{chgauss}) simplifies to 
\begin{align}\label{chgauss2}
&\CC_{H(x)}(t_1, t_2) =\sum_{k=0}^{\infty}\frac{\mathscr{G}^{(\alpha)}_{k}(\sigma_1,\gamma_1)\mathscr{G}^{(\alpha)}_{k}(\sigma_2,\gamma_2)}{k!}\,\CC_{s}^k(t_1,t_2), 
\end{align} 
which is independent of the cross-correlation of the processes $\r_1(t), \r_2(t)$. Note, however,  that the coefficients in (\ref{chgauss3}) are functions of the variances of the two processes; thus,  the mapping  $\CC_{H(x)} \mapsto \CC_s$  induced by (\ref{chgauss3}) depends on the variances, $\gamma_1^2, \gamma_2^2$,  of the corrupting processes $\r_1(t),\r_2(t)$. However, it turns out that for the ideal limiter (i.e, $H_\alpha$ with $\alpha=0$) the dependence on $\gamma_1,\gamma_2$ in (\ref{chgauss3}) can be approximately factored out as follows (see figure \ref{error})
\begin{align}\label{chgauss4}
&\CC_{H(x)}(\tau) =\frac{2}{\pi}\arcsin\Big(c(\gamma_1,\gamma_2)\,\CC_s(\tau)\Big),
\end{align}
so that the unwanted dependence on $\gamma$ when estimating $\CC_s$ from (\ref{chgauss4}) can be removed by the following normalization 
\begin{align}\label{chgauss5}
&\CC_s(\tau) = \sin\Big(\textstyle\frac{\pi}{2}\CC_{H(x)}(\tau)\Big)/\sin(\frac{\pi}{2} \CC_{H(x)}(0)); 
\end{align}
the high accuracy of this approximation is illustrated in figure \ref{error}. Additional illustration  of the high accuracy of the estimation of the correlation function $\CC_s$ from the corrupted input is shown in figure~\ref{exampl} where, for simplicity, 
 both processes are assumed identical and stationary so that the correlation of the modulated signal (\ref{chgauss4}) reduces to 
\begin{align}\label{chgauss3}
&\CC_{H(x)}(\tau) =\sum_{k=0}^{\infty}\frac{\Big(\mathscr{G}^{(\alpha)}_{k}(\sigma,\gamma)\Big)^2}{k!}\,\CC_{s}^k(\tau). 
\end{align}

\end{document}